\documentclass[12pt,a4paper]{article}
\usepackage[english]{babel}

\usepackage{amsmath}
\usepackage{amssymb}
\usepackage{amsthm}
\usepackage{bbm}
\usepackage[dvips]{graphicx}
\usepackage{verbatim}
\newcommand{\R}{\mathbb{R}}
\newcommand{\N}{\mathbb{N}}
\newcommand{\Z}{\mathbb{Z}}

\newcommand{\E}{\mathcal{E}}
\newcommand{\B}{\mathcal{B}}
\newcommand{\G}{\mathcal{G}}

\newcommand{\defeq}{\stackrel{\mathrm{def}}{=}}
\newcommand{\approxlim}{\stackrel{n \to \infty}{\approx}}

\newtheorem{thrm}{Theorem}[section]
\newtheorem{dfntn}[thrm]{Definition}

\newtheorem{crllr}[thrm]{Corollary}
\newtheorem{prpstn}[thrm]{Proposition}

\numberwithin{equation}{section}
\numberwithin{figure}{section}
\numberwithin{table}{section}

\DeclareMathSymbol{\Lambda}{\mathalpha}{letters}{"03}
\DeclareMathSymbol{\Omega}{\mathalpha}{letters}{"0A}

\begin{document}
\title{Non-sequential recursive pair substitutions and numerical entropy estimates in symbolic dynamical systems} \author{Lucio M. Calcagnile\footnote{Scuola Normale Superiore, Pisa, Italy: l.calcagnile@sns.it}$,\ $ Stefano Galatolo\footnote{Dipartimento di Matematica Applicata, Universit\`a di Pisa, Italy: galatolo@mail.dm.unipi.it}$\ $ and Giulia Menconi\footnote{Istituto Nazionale di Alta Matematica, Roma, Italy}} \maketitle
%\tableofcontents

\begin{abstract}
We numerically test the method of non-sequential recursive pair substitutions to estimate the entropy of an ergodic source. We compare its performance with other classical methods to estimate the entropy (empirical frequencies, return times, Lyapunov exponent). We considered as a benchmark for the methods several systems with different statistical properties: renewal processes, dynamical systems provided and not provided with a Markov partition, slow or fast decay of correlations. Most experiments are supported by
rigorous mathematical results, which are explained in the paper.
\end{abstract}
%---------------------------------------------------------------------------------------
\section{Introduction}
We investigate a symbolic substitution method as a tool to estimate entropy of an ergodic source. The entropy we deal with is the Shannon entropy of finite-alphabet stationary stochastic processes, in particular those that can be obtained as a symbolic model of a dynamical system.

Throughout the paper, we shall refer to this method as Non-Sequential Recursive Pair Substitution (NSRPS). The idea of applying recursive pair substitutions to symbolic sequences was first proposed by Jimenez-Monta\~no, Ebeling and others (see \cite{jimenezmontano-ebeling}), but it was put into the formal context of probability theory and studied more deeply by Grassberger \cite{grassberger} in 2002 and Benedetto, Caglioti, Gabrielli \cite{bcg} in 2006.

We now briefly explain how the NSRPS method works.

Let us suppose to have a finite-state stationary source, that is a device providing infinite sequences of symbols $x_0 x_1 x_2 \ldots$ where each $x_i$ is an element of a finite alphabet $A$, in such a way that the probability of receiving a given finite string does not vary with time. Given a sequence from such a source, the NSRPS method prescribes to individuate the pair (or one of the pairs) of symbols of maximal frequency and to substitute all its non-overlapping occurrences with a new symbol $\alpha \notin A$. For example, given the sequence
\begin{equation*}
011010111011000111011010011 \ldots ,
\end{equation*}
taken from a source $\mu$ for which $\mu (01)$ is the highest among the probabilities of symbol pairs, we substitute the pair $01$ with the new symbol $2$, thus obtaining
\begin{equation*}
2122112100211212021 \ldots .
\end{equation*}
In the case the pair to substitute is made up of two equal symbols, not \emph{all} the occurrences are to be substituted, but only the non-overlapping ones. For example, given the sequence
\begin{equation*}
00110100001010001000001100001 \ldots ,
\end{equation*}
we substitute the pair $00$, obtaining
\begin{equation*}
211012210120122011221 \ldots .
\end{equation*}

Starting from a source $\mu$ with alphabet $A = \{ 0,1 \}$, after the first substitution we shall have a new source with alphabet $A_1 = \{ 0,1,2 \}$ and a measure $\mu_1$ on the finite strings inherited from $\mu$. We can then go on repeating the steps, introducing new symbols $3,4,\ldots$ and obtaining new sources $\mu_2,\mu_3,\ldots$.

The main theorem about the NSRPS method (Theorem \ref{main_theorem}) says that the entropy $h$ of an ergodic source $\mu$, which is defined by
\begin{equation*}
h (\mu) = \lim_{k \to \infty} - \frac{1}{k} \sum_{\textrm{length}(\underline{x}) = k} \mu (\underline{x}) \log_2 \mu (\underline{x}),
\end{equation*}
can be calculated, in the limit for the number $N$ of substitutions which approaches infinity, knowing only the probabilities according to $\mu_N$ of the individual symbols and of the pairs in the new sources, after many substitutions. We remark that the hypotheses of substituting at each step one of the pairs with the maximum probability is a sufficient but not necessary condition for the conclusion of the main theorem \ref{main_theorem} to hold (see \cite{bcg}).

Numerical results about the use of this method for the estimation of the entropy of the english language were sketched in \cite{grassberger}. Here we show a first systematic comparation of this method with other classical ones, by performing several experiments on artificial sequences. We will mainly use symbolic sequences constructed by dynamical systems.

The use of symbolic mo\-dels of dynamical systems as a benchmark for
this kind of study is motivated by the following two important
features:
\begin{itemize}
\item dynamical systems can produce strings with many kinds of
  nontrivial statistical features (slow decay of correlations, no
  Markov structure, and so on...)
\item the dynamical/geometrical properties of the system under
  consideration often allow the
  entropy of the system to be estimated (sometime rigorously calculated) by some other method (Lyapunov exponents and
  geometrical properties of the invariant measure e.g.) whose results
  can be compared with the estimation done by symbolic methods.
\end{itemize}

%The use of symbolic models of dynamical systems as a benchmark for this kind of study is motivated by the fact that, while these systems produce symbolic processes with nontrivial statistical features (slow decay of correlations, no Markov structure e.g.) we can compare the results which are obtained by symbolic methods, as the NSRPS, with other methods of completely different nature which are proved to converge to the same result (for example the estimation of Lyapunov exponents).

In order to judge the precision and the speed of the entropy estimating algorithm suggested by the NSRPS method, we shall compare it with other three much used entropy estimating methods. Two of them apply to symbolic sequences. They are the \emph{empirical frequencies} method and the \emph{return times} method. Finally,  in the case of ergodic transformations, we calculate the Lyapunov exponent which converges very fast and will be considered as a reference value for the entropy. The use of these numerical estimators will be supported by rigorous mathematical results, which will be explained in the paper.

In section \ref{sec:NSRPS} we formally present the NSRPS method and state the main theorem about it. In section \ref{sec:symbolic_dynamics} we recall some basic notions of symbolic dynamics. In section \ref{sec:other_methods} we give a review of rigorous results supporting the estimation of entropy by the other methods we chose: empirical frequencies, return times and Lyapunov exponent. In section \ref{secsimul} we discuss the details of the implementation of the above methods and the reasons of some arbitrary choice we could not avoid. In section \ref{sec:experiments} we present the experimental results, with some tables and figures.

%---------------------------------------------------------------------------------------
\section{Non-Sequential Recursive Pair Substitutions (NSRPS)} \label{sec:NSRPS}
In this section we briefly recall from \cite{bcg} definitions and main results on the NSRPS method. We introduce the terms and the notations which are fundamental to state the main theorem \ref{main_theorem}. We omit all the technical details and the proofs, which the interested reader can find in \cite{bcg}.

We recall from the introduction that the method we study is applied to symbolic sequences which are supposed to come from a finite-state stationary source.

Let us call our finite alphabet $A$ and denote with $A^\ast = \cup_{k=1}^{\infty} A^k$ the collection of all finite words in the alphabet $A$. A word $\underline{w} \in A^\ast$ has length $|\underline{w}|$ and, if $|\underline{w}| = k$, it will also be indicated with $w_1^k = w_1 \ldots w_k$. 

Let $x,y \in A$, $\alpha \notin A$ and $A_1 = A \cup \{ \alpha \}$.

\begin{dfntn}
\item A pair substitution is a function $G = G_{xy}^\alpha : A^\ast \to A_1^\ast$ which is defined by recursively substituting all the non-overlapping occurrences of the pair $xy$. More precisely, $G \underline{w}$ is defined substituting in $\underline{w}$ the first occurrence from left of $xy$ with $\alpha$ and repeating this procedure to the end of the sequence.
\end{dfntn}
We consider again the example sketched in the introduction and show some general notation. Given the sequence
\begin{equation*}
\underline{w} = 011010111011000111011010011 \in \{ 0,1 \}^\ast,
\end{equation*}
performing the substitution $01 \mapsto 2$ leads to
\begin{equation*}
G_{01}^2 (\underline{w}) = 2122112100211212021 \in \{ 0,1,2 \}^\ast.
\end{equation*}

We indicate with $\E (A)$ the set of all the stationary ergodic measures on $A^\Z$, the only ones we shall deal with. If $\mu \in \E (A)$ and $\underline{w} \in A^\ast$, we shall use the notation $\mu (\underline{w})$ to indicate the $\mu$-measure of the cylinder set $[w_1,\ldots,w_k] = \cap_{i=1}^k \{ X_i = w_i \}$, where the $X_i$'s are the random variables which describe the stochastic process. 

The map $G = G_{xy}^\alpha$ naturally induces a map $\G = \G_{xy}^\alpha : \E (A) \to \E (A_1)$, as the following theorem shows. We indicate with $\sharp \{ \underline{s} \subseteq \underline{r} \}$ the number of occurrences of a subword $\underline{s}$ in a word $\underline{r}$.
\begin{thrm}
If $\mu \in \E (A)$ and $\underline{s} \in A_1^\ast$, then the limit
\begin{equation*}
\G \mu (\underline{s}) = \lim_{n \to \infty} \frac{\sharp \{ \underline{s} \subseteq G(w_1^n) \}}{|G (w_1^n)|}
\end{equation*}
exists and is constant $\mu$ almost everywhere in $\underline{w}$. Furthermore, the values $\{ \G \mu (\underline{s}) \}_{\underline{s} \in A_1^\ast}$ are the marginals of an ergodic measure on $A_1^\Z$.
\end{thrm}

It is obvious that a pair substitution shortens the sequence it is applied to. The following proposition gives an average quantification of this shortening.
\begin{prpstn}\label{Z_xy}
If $x \neq y$ then
\begin{equation}\label{eqn:Z_xy}
Z_{xy}^\mu \defeq \lim_{n \to \infty} \frac{n}{|G (w_1^n)|} = \frac{1}{1 - \mu (xy)} \quad (\mu \textrm{ a. e. in } \underline{w}).
\end{equation}
If $x = y$ then
\begin{equation}\label{eqn:Z_xx}
Z_{xx}^\mu \defeq \lim_{n \to \infty} \frac{n}{|G (w_1^n)|} = \frac{1}{1 - \sum_{k = 2}^\infty (-1)^k \mu (\underline{x}^k)} \quad (\mu \textrm{ a. e. in } \underline{w}),
\end{equation}
where $\underline{x}^k$ is the string made up of $k$ symbols $x$.
\end{prpstn}

We now recall the definition of entropy of a process. 

Given $\mu \in \E (A)$ and $n \geq 1$, the quantity
\begin{equation*}
H_n (\mu) = - \sum_{|\underline{w}| = n} \mu (\underline{w}) \log_2 \mu (\underline{w})
\end{equation*}
is the \emph{$n$-th order entropy}. 

The \emph{$n$-th order conditional entropy} is defined as 
\begin{equation*}
h_n (\mu) = H_{n + 1} (\mu) - H_n (\mu).
\end{equation*}
It can be shown (see \cite{shields}) that the quantities $h_n (\mu)$ and $H_n (\mu) / n$ converge to the same value, which is the \emph{Shannon entropy of the process $\mu$}: 
\begin{equation}\label{defshannon}
h (\mu) = \lim_{n \to \infty} h_n (\mu) = \lim_{n \to \infty} H_n (\mu) / n.
\end{equation}

%--------------------------------------------
\subsection{The main theorem}\label{sec:main_theorem}
Intuitively, after a pair substitution the information is more
 concentrated, with respect to the original
 sequence. 

 After several substitutions, the most important blocks (the most
  frequent ones) are concentrated into
symbols and the value of the entropy can be calculated by applying the standard
formula with short blocks ($H_{k}$ with small $k$).

 This can be formulated in precise terms (see \cite{bcg}, Theorem 3.2 and Corollary \ref{corollario}) and suggests that a sequence of substitutions might asymptotically transfer \emph{all} the information to the distribution of the pairs and individual symbols. This is precisely the content of the main theorem.

To state it, we define the following objects:
\begin{itemize}
\item[-] {the alphabets $A_N = A_{N-1} \cup \{\alpha_N\}$ where $\alpha_N \notin A_{N-1}$ and $A_0 = A$;}
\item[-] {the maps $G_N = G_{x_N y_N}^{\alpha_N} : A_{N-1}^\ast \to A_N^\ast$, where $x_N,y_N \in A_{N-1}$;}
\item[-] {the maps between measures $\G_N = \G_{x_N y_N}^{\alpha_N}$;}
\item[-] {the measures $\mu_N = \G_N \mu_{N-1}$, with $\mu_0 = \mu$;}
\item[-] {the quantities $Z_N = Z_{x_N y_N}^{\mu_{N-1}}$ and $\overline{Z}_N = Z_N \ldots Z_1$.}
\end{itemize}

\begin{thrm}\label{main_theorem}[\cite{bcg}, Theorem 3.2]
If
\begin{equation*}
\lim_{N \to \infty} \overline{Z}_N = + \infty
\end{equation*}
then
\begin{equation}
h (\mu) = \lim_{N \to \infty} \frac{h_1 (\mu_N)}{\overline{Z}_N}.
\end{equation}
\end{thrm}

\begin{thrm}\label{suff_cond}
If at each step $N$ the pair $x_N y_N$ is a pair with the maximum frequency among all the pairs of symbols of $A_{N-1}$, then
\begin{equation*}
\lim_{N \to \infty} \overline{Z}_N = + \infty.
\end{equation*}
\end{thrm}

Theorems \ref{main_theorem} and \ref{suff_cond} combined guarantee
that, by performing at each step the substitution of a pair with
maximum probability, the entropy of the original ergodic process is
approximated by the $1$-st order conditional entropy, which takes into
consideration only the distribution of the single symbols and of the
pairs of symbols. In this sense, through this method ``the ergodic
process becomes 1-Markov in the limit''.

In practical utilizations of the above theorem we have access to the statistical properties of the source by measuring the empirical frequency of digit sequences in the experimental data we have. Given a sequence $x_{1}x_{2}\ldots x_{n}$, the empirical distribution of the (overlapping) $k$-blocks $a_{1}^{k}$  is defined naturally by
\begin{equation}\label{empirical_distribution}
p_{k}(a_{1}^{k}|x_{1}^{n})=\frac{\#\{i\in \lbrack
1,n-k+1]:\,x_{i}^{i+k-1}=a_{1}^{k}\}}{n-k+1}
\end{equation}
and its empirical $k$-entropy is defined by
\begin{equation*}
\tilde{H}_{k}(x_{1}^{n})=-\sum_{|\underline{w}|=k}p_{k}(\underline{w}%
|x_{1}^{n})\log _{2}p_{k}(\underline{w}|x_{1}^{n}).
\end{equation*}

Let us call $G$ the substitution operation on the maximal frequency pair
(if there are more than one string of maximal frequency, the lexicographic
order is used). By ergodicity, it is possible to rephrase the above theorem
into a statement which is more similar to what can be pratically done on
long strings coming from the source:

\begin{crllr}\label{corollario}
If $\mu $ is ergodic, for almost each $\omega \in A^{\mathbb{N}}$ 
\begin{equation}
h(\mu )=\lim_{n\rightarrow \infty }\lim_{l\rightarrow \infty }\frac{\tilde{H}%
_{2}(G^{n}(\omega _{1}^{l}))-\tilde{H}_{1}(G^{n}(\omega _{1}^{l}))}{\tilde{Z}%
_{n}(\omega _{1}^{l})}
\end{equation}%
where $\tilde{Z}_{n}(\omega _{1}^{l})=\frac{l}{|G^{n}(\omega _{1}^{l})|}$ is
the shortening rate after $n$ substitutions.
\end{crllr}

\begin{proof}

 Let $\omega $ be a typical realization of the system. Since the system is
ergodic  $\lim_{l\rightarrow \infty }\tilde{H}_{k}(G^{n}(\omega
_{1}^{l}))=H_{k}({\mu }_n )$, hence $\lim_{l\rightarrow \infty }\tilde{H}%
_{2}(G^{n}(\omega _{1}^{l}))-\tilde{H}_{1}(G^{n}(\omega _{1}^{l}))=h_{1}({\mu }_n
)$. Moreover, in the same way, when $n$ is fixed and $l\rightarrow \infty $, $\tilde{Z}_{n}(\omega _{1}^{l})\rightarrow \overline{Z}_{n}$ and the corollary follows from the above Theorem \ref{main_theorem}.

\end{proof}

%---------------------------------------------------------------------------------------
\section{Symbolic dynamics} \label{sec:symbolic_dynamics}
In this section we briefly recall the basic notions about symbolic
dynamics and Kolmogorov-Sinai entropy. 
We already defined the entropy of a symbolic process. Entropy may be
defined also for measure-preserving transformations. This will be done
by associating symbolic sequences with the orbits of the
transformation. 
Let us more precisely recall the
definition of Kolmogorov-Sinai entropy $h_\mu$ of a map $T:(X,\B,\mu)\rightarrow (X,\B,\mu) $ having an ergodic invariant measure $\mu $.

Let $\alpha = \{ A_1,\ldots,A_k \}$ be a finite measurable partition of $X$. Let $\Omega$ be the pro\-duct space $\{ 1,2,\ldots,k \}^\N$, so that an element of $\Omega$ is a sequence\linebreak $\omega = (\omega_n)_{n=0}^\infty$, where $\omega_n \in \{ 1,2,\ldots,k \}$ for all $n$. 

It is possible to translate in a standard way the dynamics of
$(X,\B,\mu,T)$ into the dynamics of the space $\Omega$, which is
provided with the Borel $\sigma$-algebra $\B (\Omega)$ generated by
the cylinder sets and the left shift transformation $\sigma$. Let us define a map $\phi_\alpha : (X,\B) \to (\Omega,\B (\Omega))$ by
\begin{equation*}
(\phi_\alpha (x))_n = \omega_n \quad \textrm{if } T^n x \in A_{\omega_n}.
\end{equation*}
so that the $n$-th coordinate of $\phi_\alpha (x)$ is the alphabet letter corresponding to the element of the partition $\alpha$ which $T^n x$ belongs to.

It holds $\phi_\alpha (T x) = \sigma (\phi_\alpha x)$, $\forall x \in
X$. Furthermore, the map $\phi_\alpha$ is measurable and naturally
transports the measure $\mu$ on $(\Omega,\B (\Omega))$ defined by setting for every measurable $E \subseteq \Omega$, $\nu (E) = \mu (\phi_{\alpha}^{-1} E)$.

Notice that in general the map $\phi_\alpha$ is not invertible, thus
it does not always give an isomorphism. However, if the partition
$\alpha$ is {\em generating}, that is the sets of the form $A_{i_1} \cap T^{-1} A_{i_2} \cap \ldots \cap T^{-(m-1)} A_{i_m}$ generate the $\sigma$-algebra $\B$, then the map $\phi_\alpha$ gives an isomorphism between $(X,\B,\mu,T)$ and $(\Omega,\B (\Omega),\nu,\sigma)$. 

If $\alpha$ and $\beta$ are two measurable partitions of $(X,\mu,T)$, their {\it joint partition} $\alpha\vee\beta$ is the set $\{A\cap B\ |\ A\in\alpha\, \ B\in\beta\}$. If $T$ is a measurable and non-singular function and $\alpha$ is a partition, then $T^{-1}\alpha$ is the partition defined by the subsets $\{T^{-1}A \ |\ A\in\alpha\}$.

Given the partition $\alpha = \{ A_1,\ldots,A_k \}$ we shall denote the Shannon entropy of the partition by
\begin{equation*}
H(\alpha)= -\sum_{i=1}^k \mu(A_i)\log_2(\mu(A_i))\ .
\end{equation*}

The entropy of the map $T$ with respect to the partition $\alpha$ is:
\begin{equation*}
\displaystyle{h_\mu(T,\alpha)=\lim_{n\rightarrow+\infty}\frac 1 n H\bigg(\bigvee_{i=0}^{n-1}T^{-i}\alpha\bigg) }.
\end{equation*}

The Kolmogorov-Sinai entropy of the dynamical system $(X,\mu,T)$ is 
\begin{equation*}
h_\mu(T)=\sup_\alpha h_\mu(T,\alpha),
\end{equation*}
where the supremum is taken over all the finite partitions.

There exist partitions whose entropy is the Kolmogorov-Sinai entropy of the map.
\begin{thrm}[Kolmogorov]
Consider a dynamical system $(X,\mu,T)$. If $\alpha$ is a generating partition with respect to the map $T$, then
\begin{equation*}
h_\mu(T)=h_\mu(T,\alpha).
\end{equation*}
\end{thrm}
The existence of a generating partition for a dynamical system is assured by the following theorem.
\begin{thrm}[Krieger Generator Theorem \cite{krieger}]
For an ergodic dynamical system $(X,\mu,T)$ on a Lebesgue space $X$, such that $h_\mu(T)<\infty$, there exists a finite generating partition $\alpha$.
\end{thrm}

The identification of a generating partition is generally a challenging task. In the following, we shall provide some examples of generating partitions in specific cases.

%---------------------------------------------------------------------------------------
\section{Estimating entropy from samples} \label{sec:other_methods}
When a process'invariant measure is explicitly known, we could in
principle estimate the entropy by applying the definition. On the
other hand, when we are not given explicit knowledge of the measure,
we are often not able to know exactly the entropy of the process and
the problem of entropy estimation arises. A usual approach to this
problem is considering long sample sequences, which are looked at as
parts of infinite typical sequences and thus representing the
statistical features of the system. To such samples several entropy
estimating algorithms can be applied.

We shall compare the estimating algorithm suggested by the NSRPS
method with two others, which we shall call the \emph{empirical
frequencies} (briefly, EF) method and the \emph{return times}
(briefly, RT) method. We remark that these methods can be applied
directly to the symbolic sequence without ha\-ving any other information
on the source. For the ergodic transformations of the unit interval
we shall use another estimating algorithm which does not apply to
symbolic processes: the approximation of the \emph{Lyapunov
exponent}. We remark that the estimation of entropy by this method
uses some additional information on the system (the derivative of the
map, which is calculated at each step of the dynamics, and the
dimension of the invariant measure).

 Each estimation algorithm is supported by
rigorous results, as it will be shown in the following sections and will be implemented in its simplest form.

We end remarking that, while experimental examples contained in this paper are long artificial trajectories mostly coming from dynamical systems. When working on short sequences (for instance finite realization of some biophysical process or experiment), surrogate analysis and a suitable correction of the estimator can be useful in order to take into account fluctuations of entropy or implicit bias on the chosen estimator (see e.g. \cite{montano02}, \cite{BHM}).

%--------------------------------------------
\subsection{Empirical frequencies (EF)}
To estimate entropy directly by the definition, a simple procedure consists in determining the empirical distribution $p_k$ of the overlapping $k$-blocks and taking $\frac{H_k (p_k)}{k}$ as an estimate for $h$. If $k$ is fixed and the length of the sample sequence $n$ tends to infinity, then $\frac{H_k (p_k)}{k}$ almost surely converges to $\frac{H_k (\mu_k)}{k}$, which tends to $h$ as $k \to \infty$. Theorem \ref{empirical_frequencies} below guarantees that these two limits can be performed together with $k (n) \sim \log_2 n$.

Given the sequence $x_1 x_2 \ldots x_n$, the empirical distribution %correzione
$p_{k} (\cdot | x_1^n)$ of the overlapping $k$-blocks is defined %correzione
 as in (\ref{empirical_distribution}).

\begin{thrm}\label{empirical_frequencies}
If $\mu$ is an ergodic measure of entropy $h > 0$, if $k (n) \to \infty$ as $n \to \infty$ and if $k (n) \leq \frac{\log_2 n}{h}$, then
\begin{equation*}
\lim_{n \to \infty} \frac{1}{k (n)} H_{k(n)} ( x_1^n) = h, \quad \textrm{a. s.}
\end{equation*}
\end{thrm}

For the proof and further details see \cite{shields}, Theorem II.3.5 and Remark II.3.6. We remark that the same result holds for non-overlapping distributions. The reason why we chose to consider the overlapping one is to enrich the statistic as much as possible, as it will be explained in section \ref{secsimul}.

%Theorem~II.3.5 in \cite{shields} is the non-overlapping version of %Theorem \ref{empirical_frequencies}, that is with the distributions $q_k$ of %the non-overlapping blocks in place of the $p_k$'s, where
%\begin{equation*}
%q_k (a_1^k | x_1^n) = \frac{| \{ i \in [0,m-1] : \, x_{ik+1}^{ik+k} = a_1^k \} %|}{m},
%\end{equation*}
%with $n = k m + r$, $0 \leq r < k$.
%
%When dealing with finite sequences, as we obviously do in our numerical experiments, the statistics of the process is better obtained looking at the overlapping block distribution.

%--------------------------------------------
\subsection{Return times (RT)}
Ornstein and Weiss proved an interesting result which links entropy and the so-called return times for ergodic processes. They showed in \cite{ornstein-weiss} that the logarithm of the waiting time until the first $n$ terms of a sequence $x$ occur again in $x$ is almost surely asymptotic to $n h$.

\begin{dfntn}
Given a sequence $x$ taken from an ergodic process, we define the $n$-th return time as
\begin{equation*}
R_n (x) = \min \{ m \geq 1 : \, x_{m+1}^{m+n} = x_1^n \}.
\end{equation*}
\end{dfntn}

\begin{thrm}
If $\mu$ is an ergodic process with entropy $h$, then
\begin{equation*}
\lim_{n \to \infty} \frac{1}{n} \log_2 R_n (x) = h, \quad \textrm{a. s.}
\end{equation*}
\end{thrm}

For the original proof see \cite{ornstein-weiss}, for an alternative one see \cite{shields}, Theorem II.5.1.

%--------------------------------------------
\subsection{Lyapunov exponent}
If we are interested in the estimation of the entropy of a one dimensional system a powerful tool is the Lyapunov exponent.

Let us consider a map $T:[0,1]\rightarrow [0,1]$ having an ergodic invariant measure $\mu $. We define its Lyapunov exponent by
\begin{equation*}
\lambda _{\mu}=\int_0^1 \log_2 T^{\prime}d\mu .
\end{equation*}
Under some assumptions (see below) this quantity is related to the fractal dimension $HD(\mu )$ of $\mu $ and the entropy $h_{\mu }$ of the system by the formula $HD(\mu )=\frac{h_{\mu }}{\lambda _{\mu }}$. Hence if we know $HD(\mu )$ and estimate $\lambda _{\mu }$ numerically, we obtain an estimation for $h_{\mu }$.

Let us give a precise statement for one dimensional systems (see \cite{LY85} for a generalization to multidimensional systems). A map $T:[0,1]\rightarrow \lbrack 0,1]$ is called piecewise monotonic if there is a sequence $\{Z_{i}\}_{i\in \mathbb{N}}$ of disjoint open subintervals of $[0,1]$ such that $T|_{Z_{i}}$ is strictly monotone and continuous for each $i$.

Let us consider the set $E_{Z} = \cap_{i \in \N} T^{-i} (\cup_{j \in
 \N} Z_{j})$, where all iterates of $T$ are in the open intervals.
Let $\mu$ be an invariant ergodic measure such that $\mu
 (E_{Z})=1$. Let us consider its Lyapunov exponent $\lambda _{\mu }$
 and its K-S entropy $h_{\mu }$. Let us denote by $HD(X)$ the
 Hausdorff dimension of a subset $X \subset [0,1]$. The Hausdorff
 dimension $HD(\mu )$ of a measure $\mu $, is defined as the infimum $HD(\mu)=\inf_{\mu (X)=1}(HD(X))$ of the dimension of full measure sets.

Let us consider the $p$-variation of a function $f$ :$[0,1]\rightarrow \mathbb{R}$ on a subinterval $[a,b]$ defined by:
\begin{equation*}
\textrm{var}_{[a,b]}^{p}(f)=\sup \bigg \{ \sum_{i=1}^{m}|f(x_{i-1})-f(x_{i})|^{p} \ \bigg | \ m \in \N,a\leq x_{0} < \ldots < x_{m}\leq b \bigg \}.
\end{equation*}
We say that the derivative of a piecewise monotonic map has bounded $p$-variation if there is a function $g$ such that $g(x)=0$ on $[0,1] \setminus E_{Z}$, $g=T^{\prime }$ on each $Z_{i}$ and $\textrm{var}_{[0,1]}^{p}(g)<\infty $.

\begin{thrm}[\cite{HR92}]\label{lyap}
Let $T$ be a map on $[0,1]$ with finitely many monotonic pieces and a derivative of bounded $p$-variation for some $p\geq 0$. If $\mu $ is an ergodic invariant measure with Lyapunov exponent $\lambda _{\mu }>0$, then
\begin{equation*}
HD(\mu )=\frac{h_{\mu }}{\lambda _{\mu }}.
\end{equation*}
\end{thrm}

In many of the systems we will study we have that the invariant
measure $\mu $ we are interested to consider is absolutely continuous with respect to the Lebesgue measure with a regular (bounded variation or continuous) density, hence $HD(\mu )=1$.

The Lyapunov exponent will be then numerically estimated with a Birkhoff average along a typical orbit of the system, hence giving
\begin{equation*}
h_{\mu }=\int_0^1 \log_2 T^{\prime }d\mu =\lim_{n\rightarrow \infty }\frac{\sum_{i=1}^{n}\log_2 (T^{\prime }(T^{i}(x_{0})))}{n}
\end{equation*}
for $\mu $-a.e.  $x_{0}$, by the ergodic theorem. Experimental results indicate that this limit converges very fast and gives a very good estimation for $h_{\mu }$.

%---------------------------------------------------------------------------------------
\section{Computer simulations}\label{secsimul}
Concerning the results of the computer simulations, some comments are due on the way we implemented the entropy estimating algorithms.
\begin{description}
 
\item$\bullet$ About empirical frequency estimation, in our simulations we could not consider blocks much longer than $23$ bits. This is because the algorithm takes a time which grows exponentially in the length of the blocks considered. The empirical distribution of blocks of various lengths was calculated on the entire symbolic sequence.

%Trying to apply the algorithm with $k = \lfloor \frac{1}{h} \log_2 n \rfloor$ (or $k = 1,2,\ldots$ and $n = \lceil 2^{hk} \rceil$) cannot give reliable results when the entropy $h$ of the transformation is low. In fact, the lower the entropy, the lower the ratio $\frac{n}{k}$, so that in some cases we should have considered an empirical distribution of the $k$ blocks obtained by few tens of bits. In such cases we could not take $n = \lceil 2^{hk} \rceil$ and took $n = 2^k$, obtaining a larger statistics.

%Independently of the results obtained on the given maps, all these computational constraints reduce the number of cases the method can be practically applied to.

\item$\bullet$ The return times method was performed by calculating the return times of strings long up to $\log_2 n$, where $n$ is the length of the symbolic sequences. Moreover, in order to have more reliable results, for every binary sequence we considered not only the return times of the initial strings $x_1^k$, but also of $x_2^{k+1}$, $x_3^{k+2}$, \ldots, $x_{1000}^{k+999}$, and took the average of their logarithms, hence what we measure is an average return time indicator.

\item$\bullet$ In the implementation of the NSRPS method, at every step the substitution with a new symbol of a pair with maximum probability was performed, then we calculated the conditional entropy of order $1$ and the inverse of the mean shortening $Z_N$ estimating the entropy according to %correzione
 Corollary \ref{corollario}.
% Proposition \ref{Z_xy}. At step $N$, that is after
%  having performed $N$ pair substitutions, we took $h_1 (\mu_N) /
% \overline{Z}_N$ as the entropy estimate, using corollary \ref{corollario}.

The implementation of the substitutions method did not show meaningful computational constraints, since performing a pair substitution requires a very short time. Nevertheless, there %correzione
is one algorithmic question to be answered: the identification of a stop condition.
% and the way to estimate the inverse of the mean shortening $Z$.

For the estimation of the entropy with NSRPS, at the moment we have not an analogous of Theorem \ref{empirical_frequencies}, hence we have to find how many substitutions it is convenient to made on a finite sample string. We had to understand when to stop the substitutions before the sequence becomes too short and consequently the statistics becomes too poor. We chose to stop when the following condition has occurred:

\emph{StopCond}: {the substituted pair has frequency $< 0.02$.}

The stop condition above is somewhat artificial and has no intrinsic relation with the symbolic process. In all the cases we studied we knew the true entropy or estimated it quite precisely by means of the Lyapunov exponent, so that we could understand when the approximation through the pair substitutions method was good. In all our processes, for which we took symbolic samples long 15 millions bits, it seems that few tens of pair substitutions are enough for the estimate to become more or less constant when considering the first three decimal digits. Obviously, when the process is independent or 1-Markov at most one pair substitution is needed in order to have a very precise estimate of the entropy. On the contrary, processes which have long memory properties need many pair substitutions. The stop condition we used does not take into account the memory properties of the process, so that it lets the algorithm performing unnecessary pair substitutions in low-Markov cases and stops it before useful substitutions in long-memory processes. Although a threshold lower than 0.02 in \emph{StopCond} could improve the estimates, the goal is to find some criterion, both user-independent and sequence-dependent, which determines for each case the most appropriate number of substitutions to perform.

\end{description}
%%%%%%%%end CANCELLARE???

%---------------------------------------------------------------------------------------
\section{Experiments} \label{sec:experiments}
We now describe the transformations of the unit interval generating the symbolic sequences to which we applied the entropy estimating algorithms.

%--------------------------------------------
\subsection{Maps}
We considered a few maps of the interval, to which we applied the construction explained in section \ref{sec:symbolic_dynamics} to obtain symbolic sequences.

%--------------------------------------------
\subsubsection{Piecewise expanding maps}\label{pwexp}
\begin{figure}[!h] 
\centerline{\bf PIECEWISE EXPANDING MAP}\vskip 10pt 
\begin{tabular}{cc}
\fbox{map}&\fbox{EF}\\ 
{\raggedright{\includegraphics[height=6cm,angle=270]{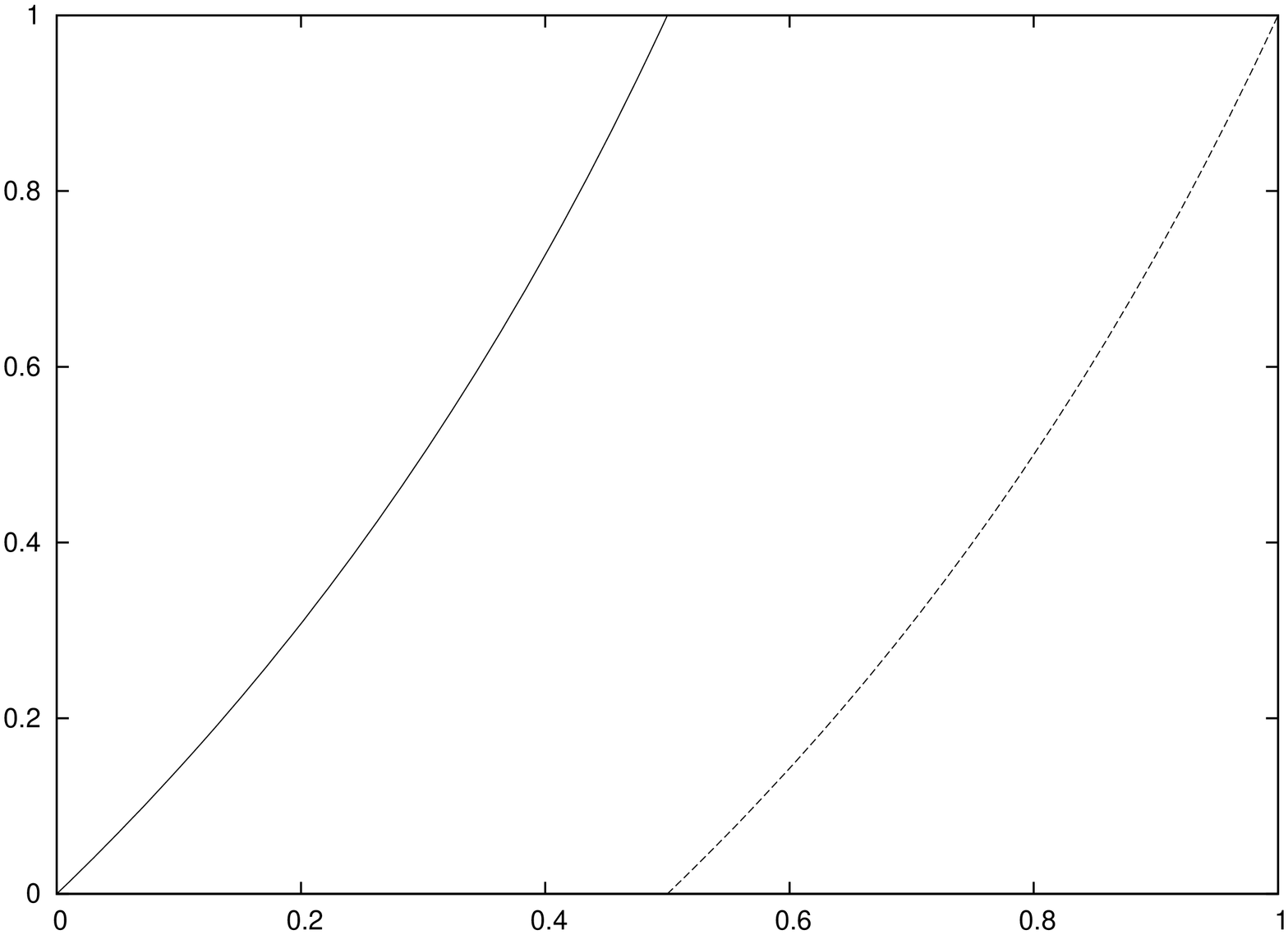}}} &
{\raggedleft{\includegraphics[height=6cm,angle=270]{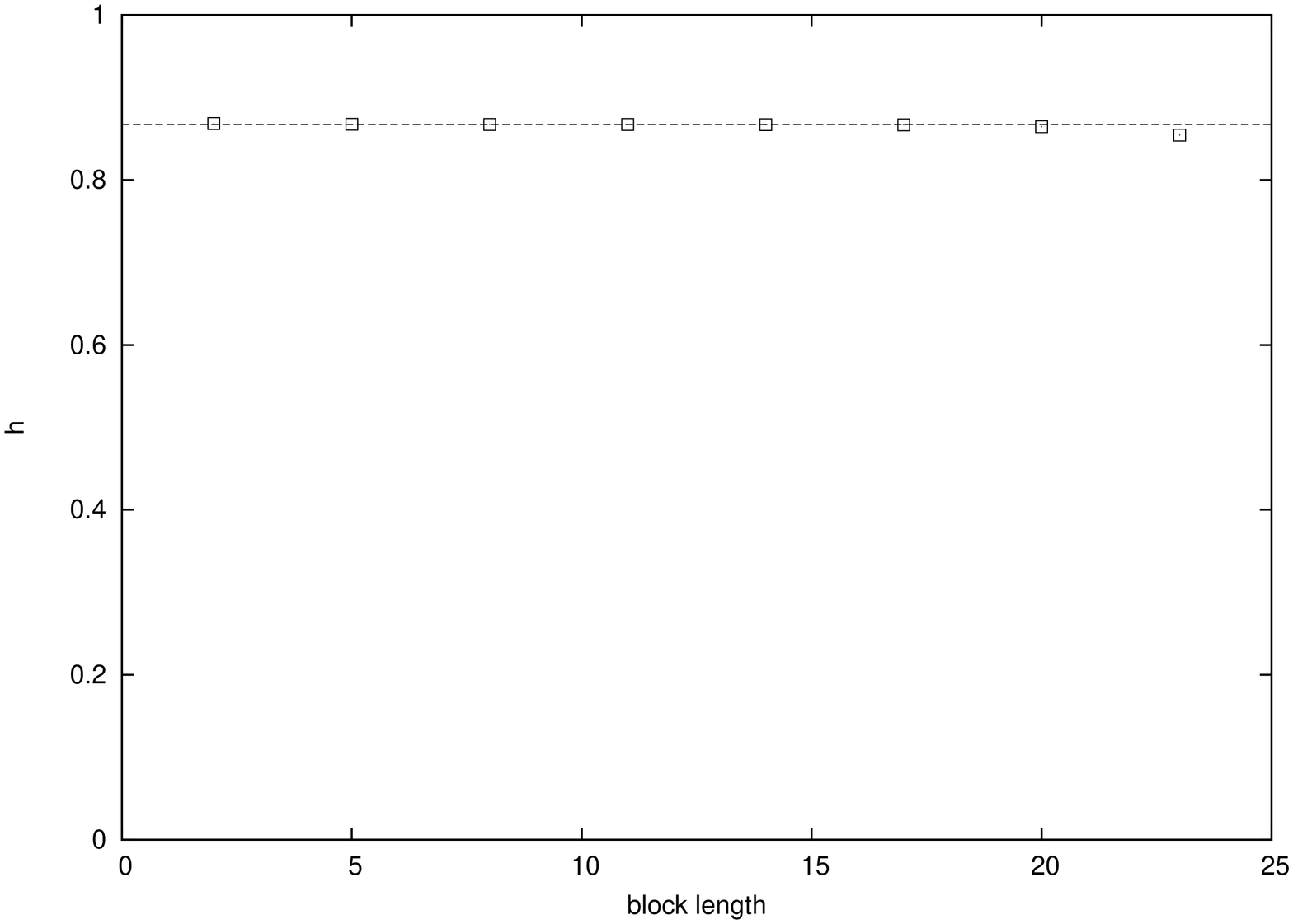}}}\\[125pt]
\fbox{RT}&\fbox{NSRPS}\\
{\raggedright{\includegraphics[height=6cm,angle=270]{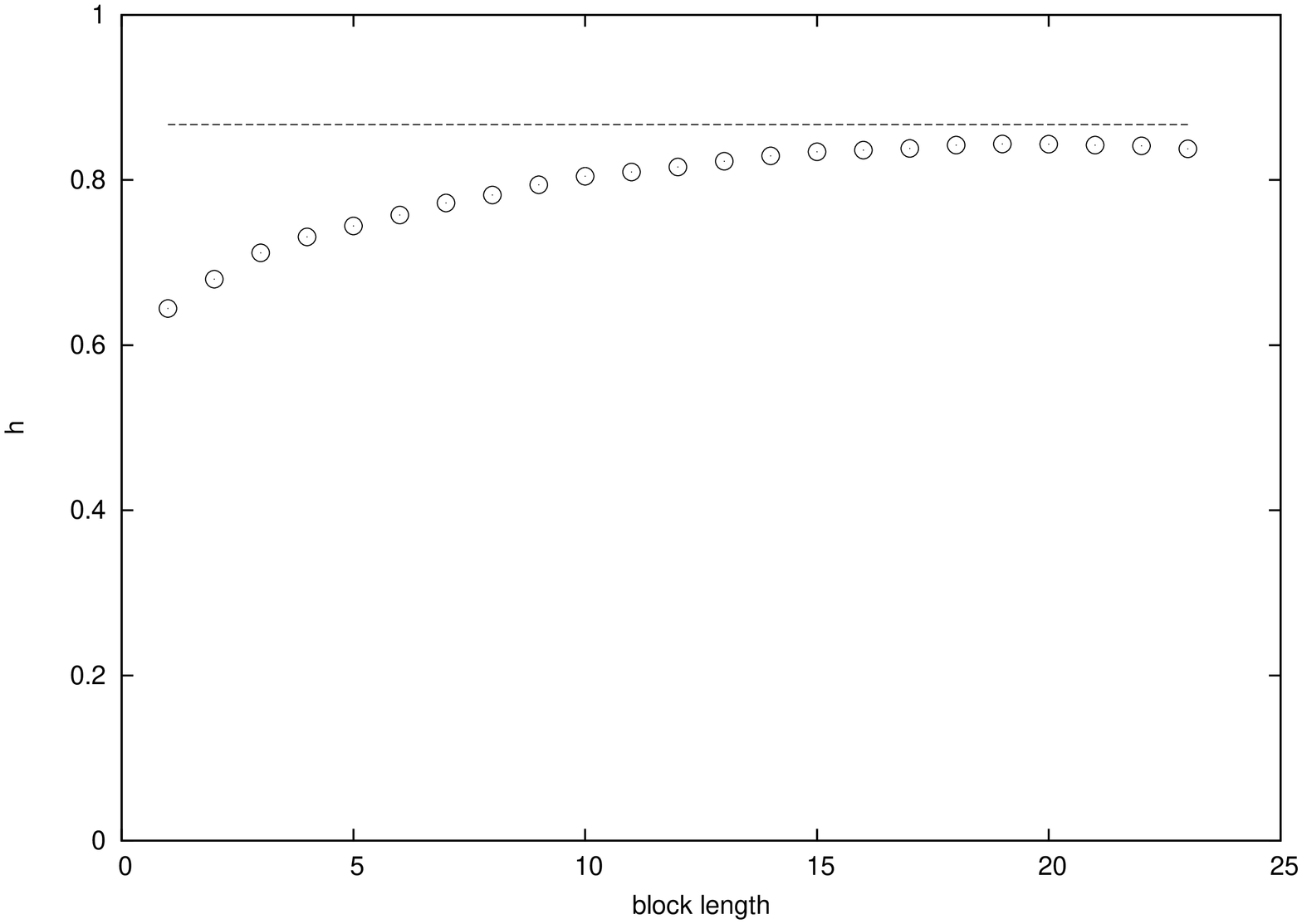}}} &
{\raggedleft{\includegraphics[height=6cm,angle=270]{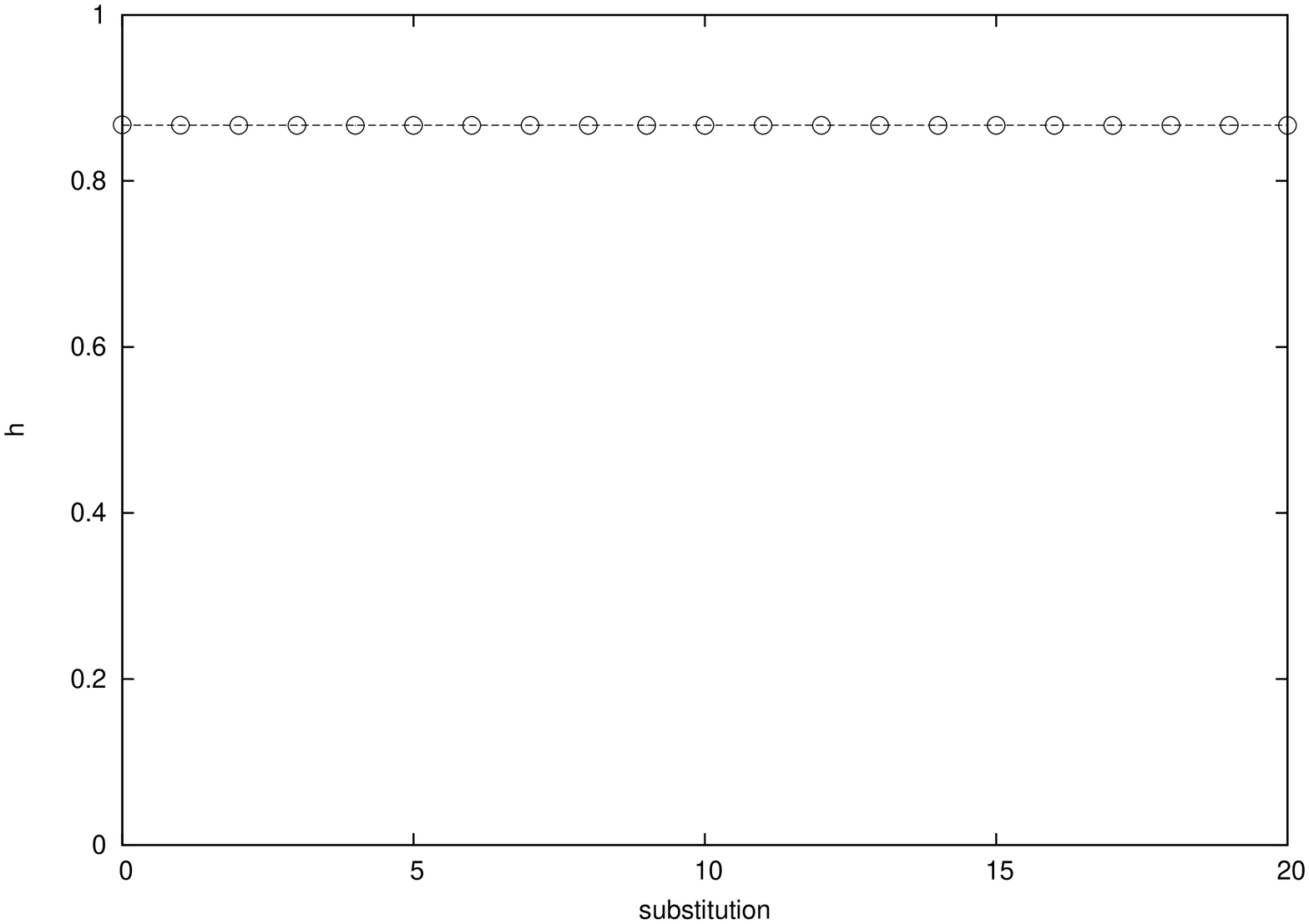}}}
\end{tabular} 
\caption{\it Piecewise expanding map $E$ and entropy estimates by means of empirical frequencies, return times and NSRPS. The straight line corresponds to the Lyapunov exponent value.}
\label{figexpand} 
\end{figure}

We considered a piecewise expanding map $E$, defined by
\begin{equation*}
E x = \left\{ \renewcommand{\arraystretch}{1.5}  % aumenta la distanza predefinita tra le righe del fattore 1.5
\begin{array}{ll}
\frac{4 x}{3 - 2 x} & \textrm{if } x \in [0,\frac{1}{2}[\\
\frac{2 x - 1}{2 - x} & \textrm{if } x \in [\frac{1}{2},1]
\end{array} \right. ,
\end{equation*}
which is discontinuous in $\frac{1}{2}$ and has two surjective branches (see Figure \ref{figexpand}). It holds $E' (x) > k$ for all $x$, where $k > 1$ is a constant. As it is well known (see e.~g.~\cite{viana}), a map of this kind has a unique absolutely continuous invariant measure with dimension $1$. Moreover, Theorem \ref{lyap} applies and we can estimate the entropy by the Lyapunov exponent. A generating partition for $E$ is $\{ [0,\frac{1}{2}[,[\frac{1}{2},1] \}$ (see \cite{buzzi}, Exercise 3.4).

We show the results of the entropy estimates in Table \ref{table-expanding} and Figure \ref{figexpand}.

% correzione (alcuni valori della tabella)
\begin{table}[!h]
\begin{equation*}
\renewcommand{\arraystretch}{1.5}  % aumenta la distanza predefinita tra le righe del fattore 1.5
\begin{array}{|l|r|r|r|r|}
\hline
\textrm{map}  & h_{\textrm{Lyap}} & h_{\textrm{EF}} & h_{\textrm{RT}} & h_{\textrm{NSRPS}} \ (N_{\textrm{sub}})\\ \hline
E             &            0.8673 &     0.865  &           0.838 &              0.867 \               (17)\\ \hline
\end{array}
\end{equation*}
\caption{\it Entropy estimates for the piecewise expanding map $E$. The values $h_{\textrm{Lyap}}$, $h_{\textrm{EF}}$, $h_{\textrm{RT}}$ and $h_{\textrm{NSRPS}}$ are the entropy estimates as Lyapunov exponent or by empirical frequencies, return times, NSRPS, respectively. $N_{\textrm{sub}}$ is the number of pair substitutions executed when the stop condition \emph{StopCond} occurs. }
\label{table-expanding}
\end{table}

The NSRPS method gives the best estimate. Though, the substitutions themselves have no particular role, since the map seems to be $1$-Markov (the first value calculated with the substitutions algorithm is already very close to the true entropy).

%--------------------------------------------
\subsubsection{Lorenz-like maps}
\begin{figure}[!h] 
\centerline{\bf LORENZ-LIKE MAP} \vskip 10pt 
\begin{tabular}{cc}
\fbox{map}&\fbox{EF}\\ 
{\raggedright{\includegraphics[height=6cm,angle=270]{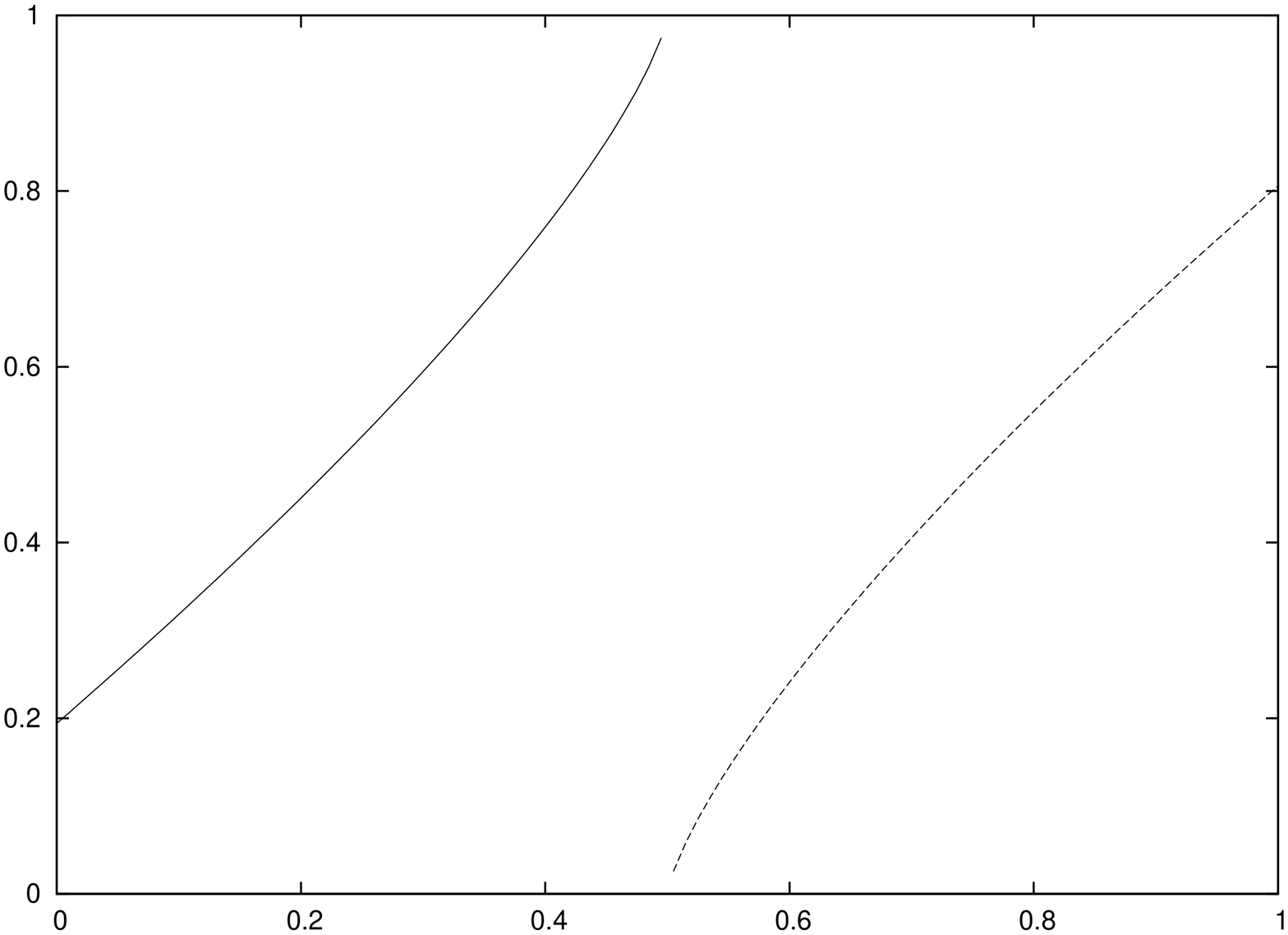}}} &
{\raggedleft{\includegraphics[height=6cm,angle=270]{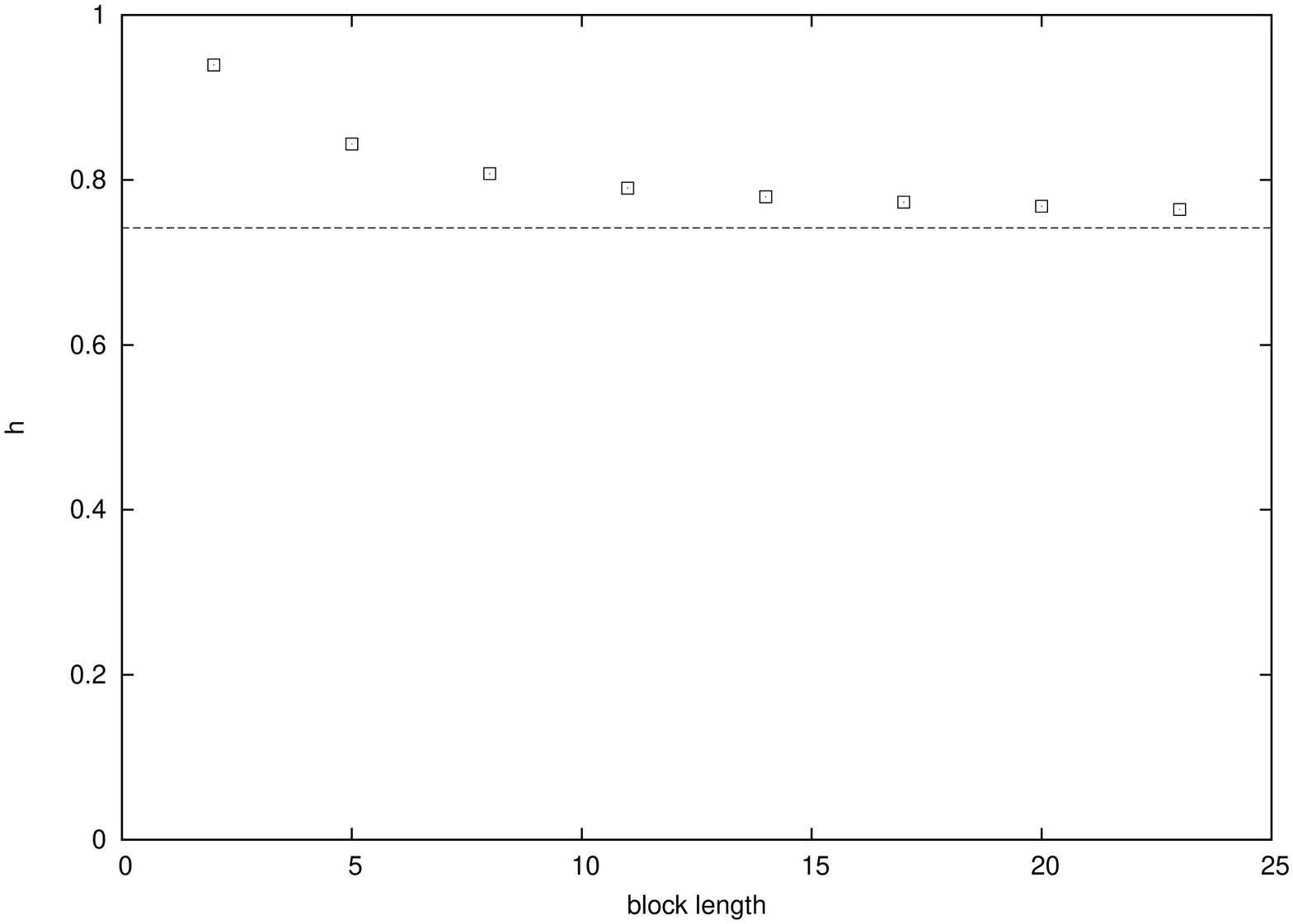}}}\\[125pt]
\fbox{RT}&\fbox{NSRPS}\\
{\raggedright{\includegraphics[height=6cm,angle=270]{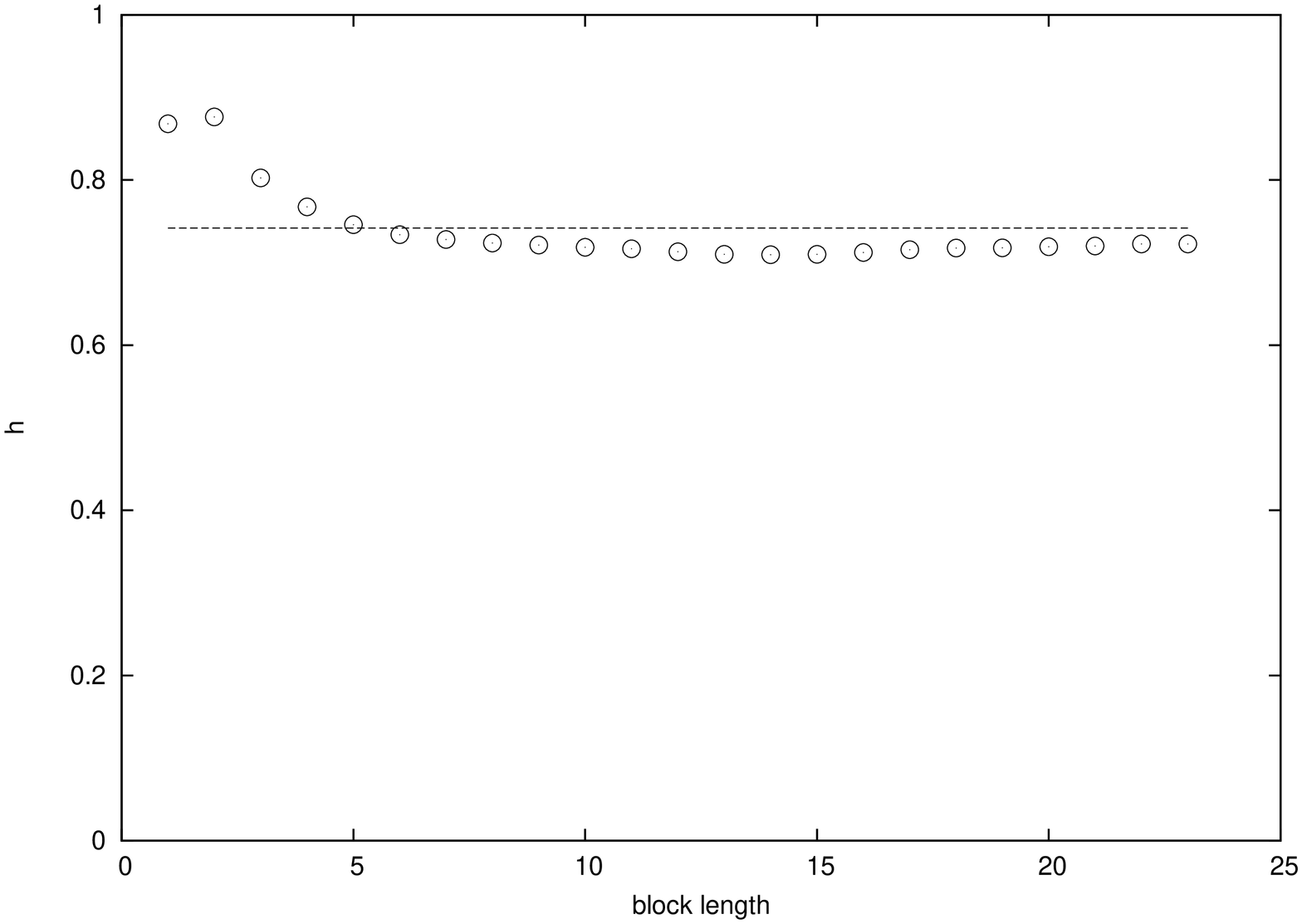}}} &
{\raggedleft{\includegraphics[height=6cm,angle=270]{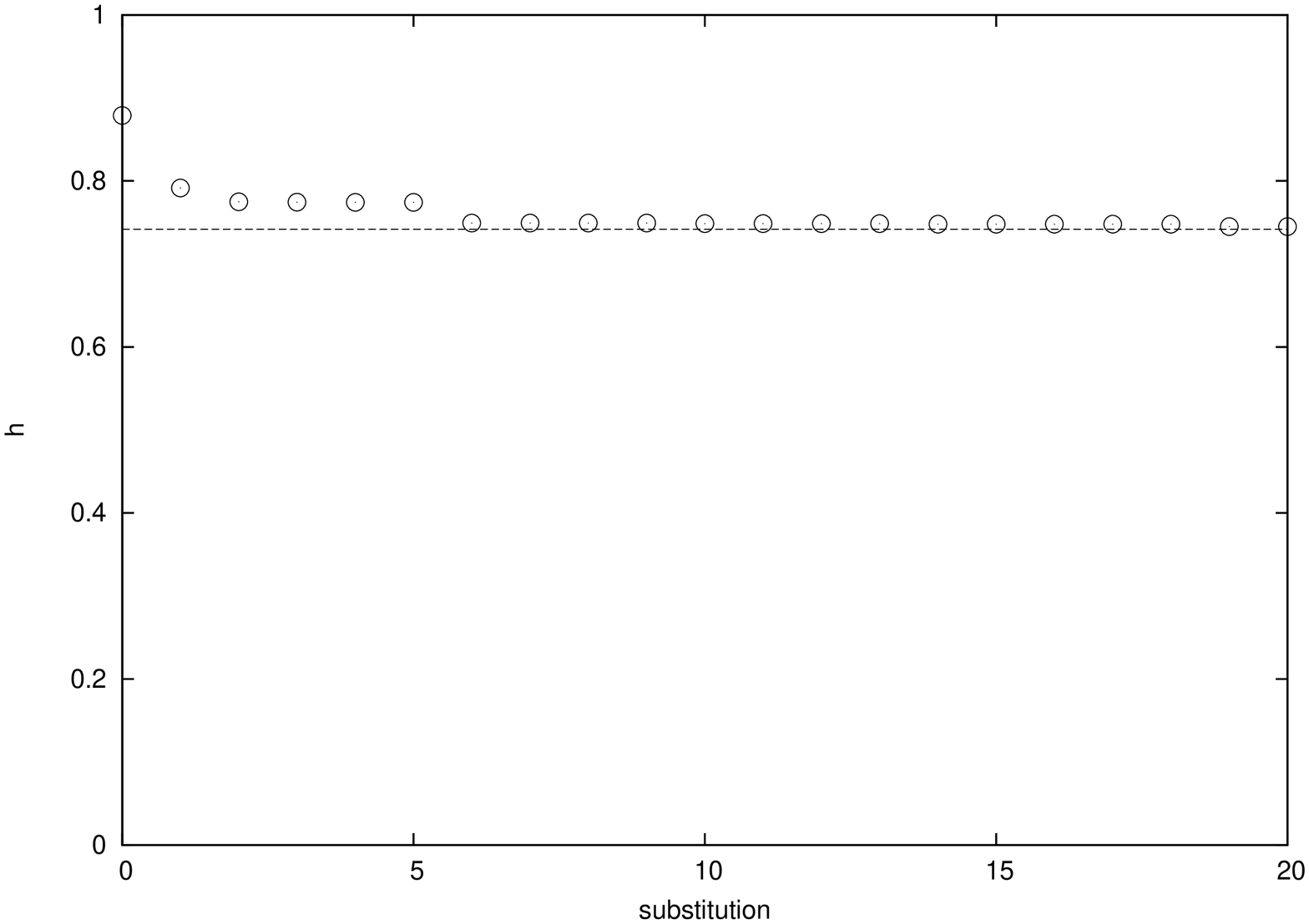}}}
\end{tabular} 
\caption{\it Lorenz-like map $L$ and entropy estimates by means of empirical frequencies, return times and NSRPS. The straight line corresponds to the Lyapunov exponent value.}
\label{figlorenz}
\end{figure}

Another example of map with two non-surjective branches is a Lorenz-like map (similar maps are involved in the study of the famous Lorenz system) defined by
\begin{equation*}
L x = \left\{ \renewcommand{\arraystretch}{1.5}  % aumenta la distanza predefinita tra le righe del fattore 1.5
\begin{array}{ll}
1 - \left( \frac{- 6 x + 3}{4} \right) ^{\frac{3}{4}} & \textrm{if } x \in [0,\frac{1}{2}[\\
\left( \frac{6 x - 3}{4} \right) ^{\frac{3}{4}} & \textrm{if } x \in [\frac{1}{2},1]
\end{array} \right. .
\end{equation*}
The derivative of $L$ is uniformly greater than $1$ for all $x \in [0,1] \setminus \{ \frac{1}{2} \}$ and $L' \big( \frac{1}{2} ^\pm \big) = + \infty$ (see Figure \ref{figlorenz}).

As for the previous piecewise expanding map $E$, the Lorenz-like map $L$ has a unique absolutely continuous invariant measure with dimension 1 (see \cite{viana}). Theorem \ref{lyap} does not apply in this case because the derivative is not bounded and hence has not $p$-bounded variation. However the usual relation between entropy and Lyapunov exponent holds and can be recovered by \cite{St}. Moreover, the natural partition $\{ [0,\frac{1}{2}[,[\frac{1}{2},1] \}$ is generating (see again \cite{buzzi}).

In Table \ref{table-lorenzlike} and Figure \ref{figlorenz} the results obtained for the map $L$ are shown.

% correzione (alcuni valori della tabella)
\begin{table}[!h]
\begin{equation*}
\renewcommand{\arraystretch}{1.5}  % aumenta la distanza predefinita tra le righe del fattore 1.5
\begin{array}{|l|r|r|r|r|}
\hline
\textrm{map}  & h_{\textrm{Lyap}} & h_{\textrm{EF}} & h_{\textrm{RT}} & h_{\textrm{NSRPS}} \ (N_{\textrm{sub}})\\ \hline
L             &            0.7419 &    0.764  &           0.723 &              0.748 \               (17)\\ \hline
\end{array}
\end{equation*}
\caption{\it Entropy estimates for the Lorenz-like map $L$. $N_{\textrm{sub}}$ is the number of pair substitutions executed when the stop condition \emph{StopCond} occurs.}
\label{table-lorenzlike}
\end{table}

The Lorenz-like map $L$ appears not to be $1$-Markov. In fact, from the plot relative to NSRPS in Figure \ref{figlorenz} it can be noticed that the best value is not the first estimated, that is simply the 1-st order conditional entropy $h_1$. Instead, there are pair substitutions that significantly improve the approximation of the entropy. These substitutions are those which condense more information than others. Furthermore, this is one of the cases in which a few more pair substitutions after condition \emph{StopCond} occurs give a better estimate.

%--------------------------------------------
\subsubsection{Logistic maps}
The logistic maps are of the form
\begin{equation*}
\Lambda_\lambda x = \lambda x (1-x), \quad 1\leq\lambda\leq 4
\end{equation*}

\begin{figure}[!h]
\centerline{\bf LOGISTIC MAP} \vskip 10pt
\begin{tabular}{cc}
\fbox{map}&\fbox{EF}\\
{\raggedright{\includegraphics[height=6cm,angle=270]{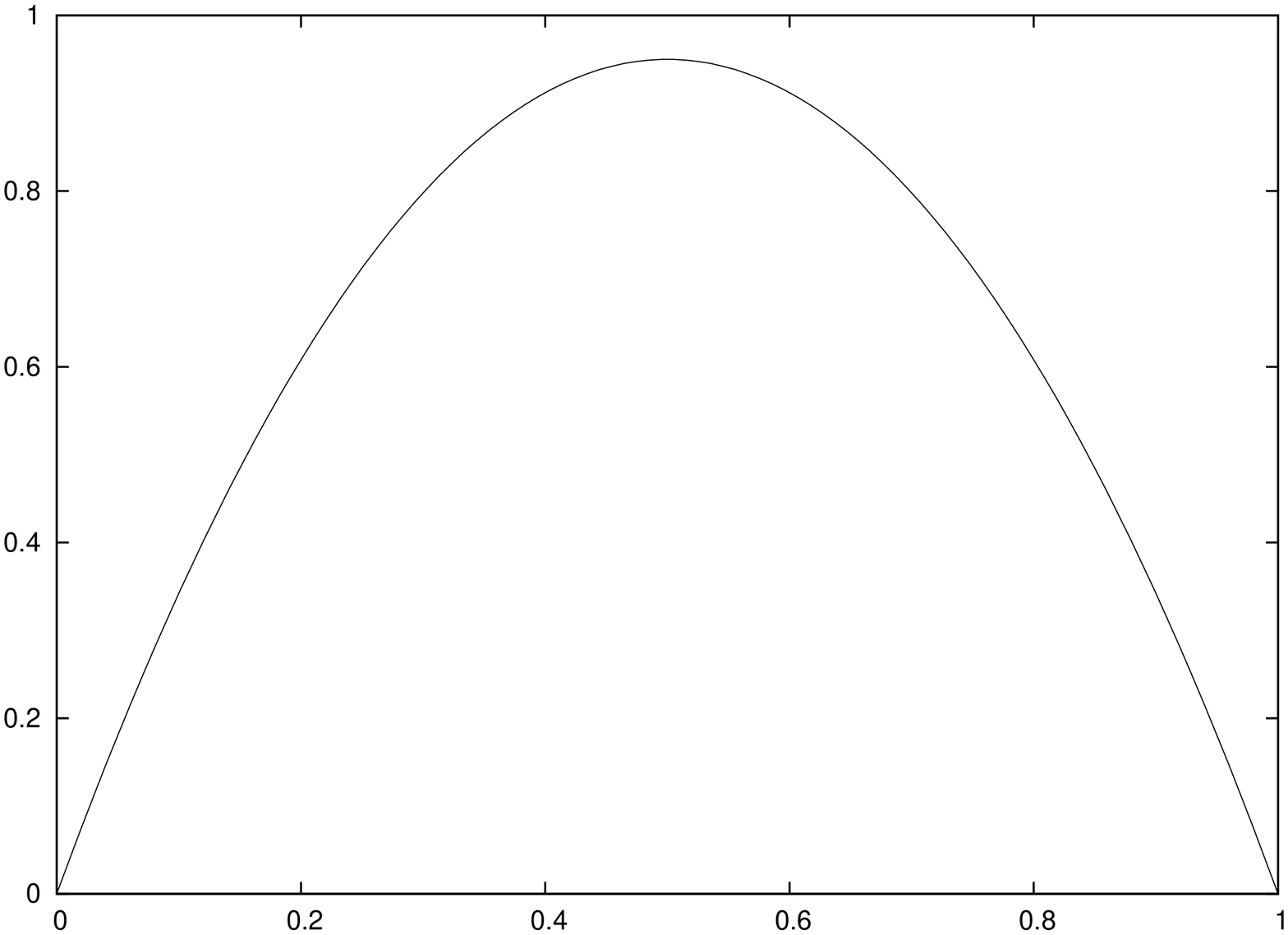}}} &
{\raggedleft{\includegraphics[height=6cm,angle=270]{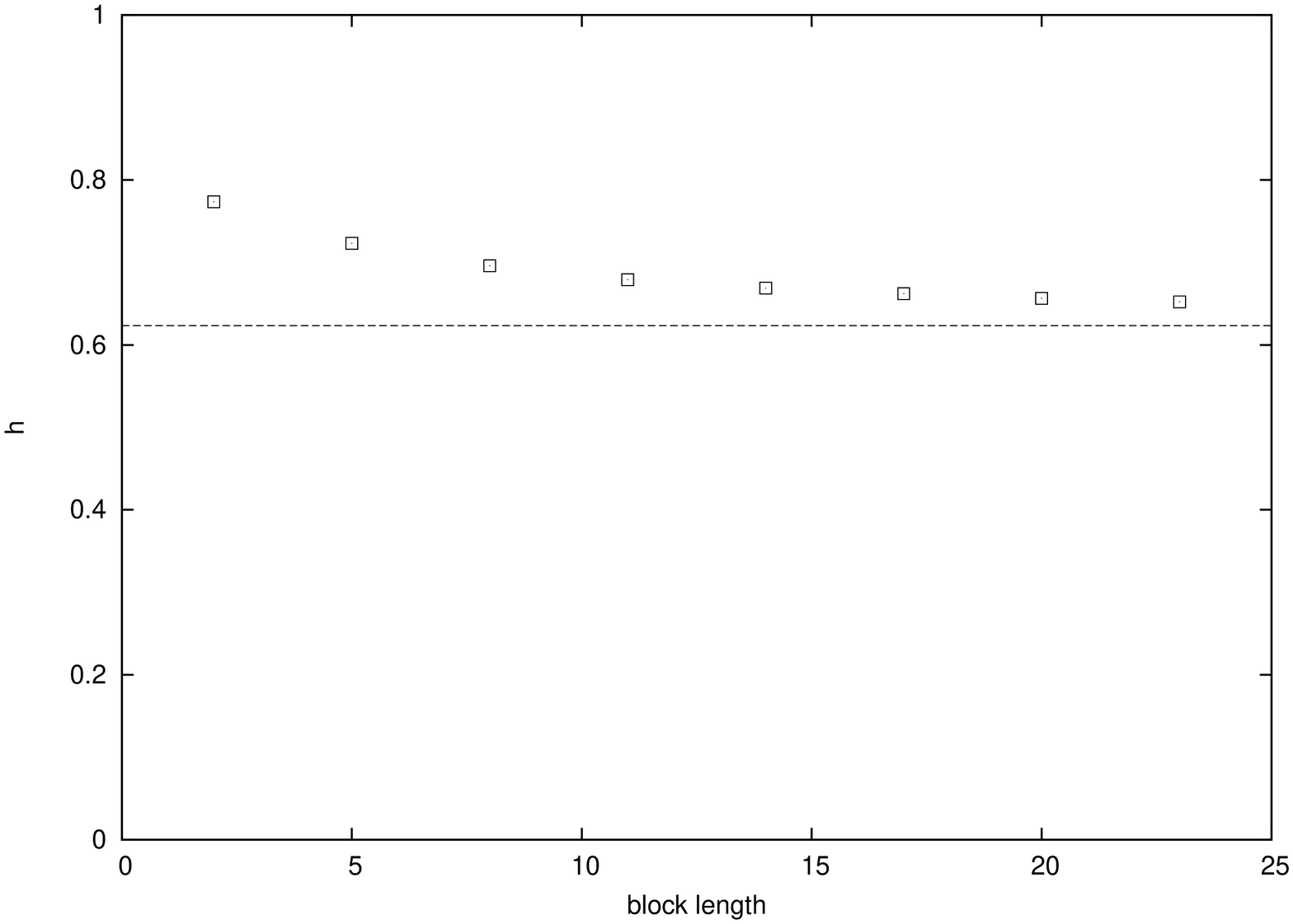}}}\\[125pt]
\fbox{RT}&\fbox{NSRPS}\\
{\raggedright{\includegraphics[height=6cm,angle=270]{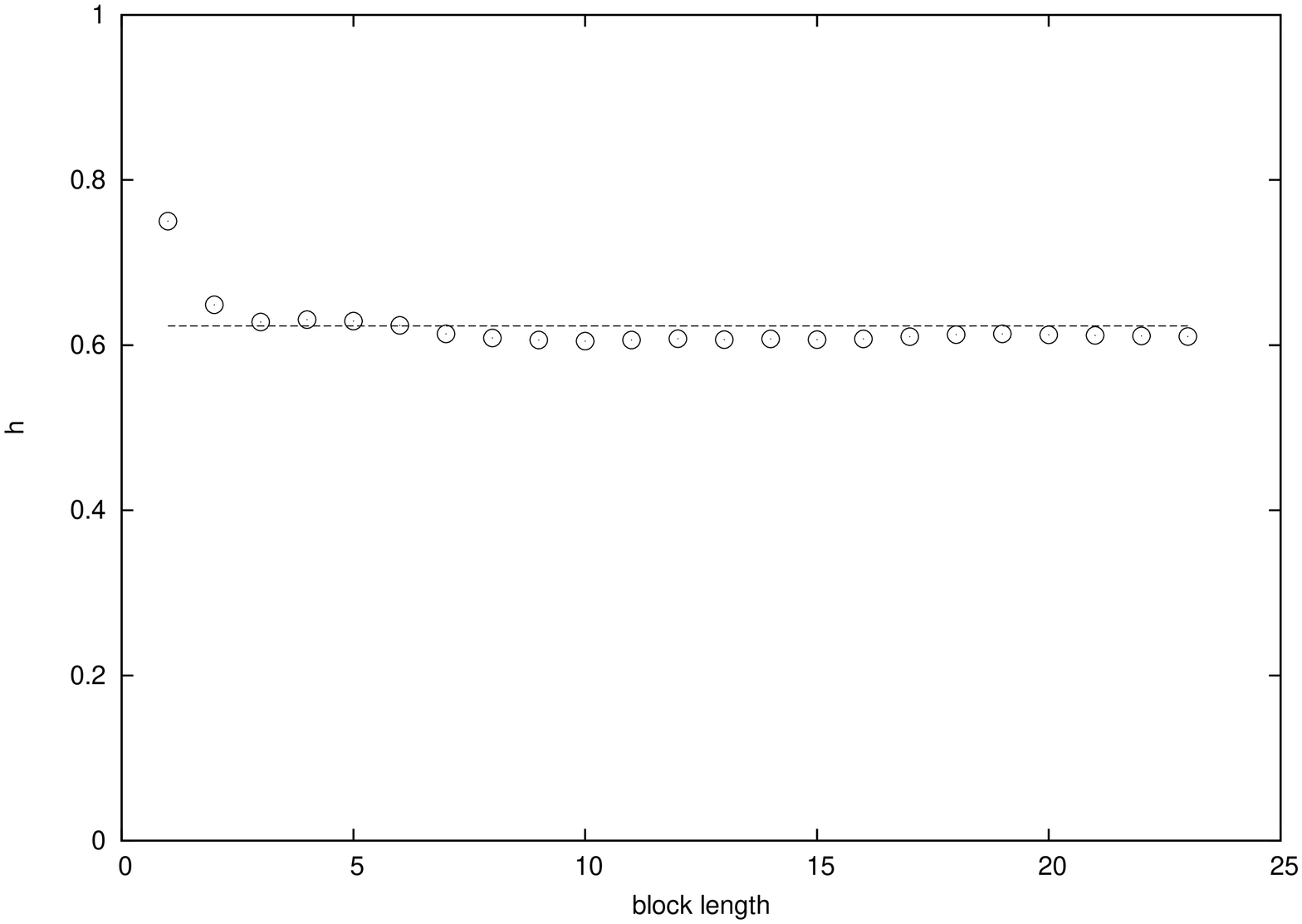}}} &
{\raggedleft{\includegraphics[height=6cm,angle=270]{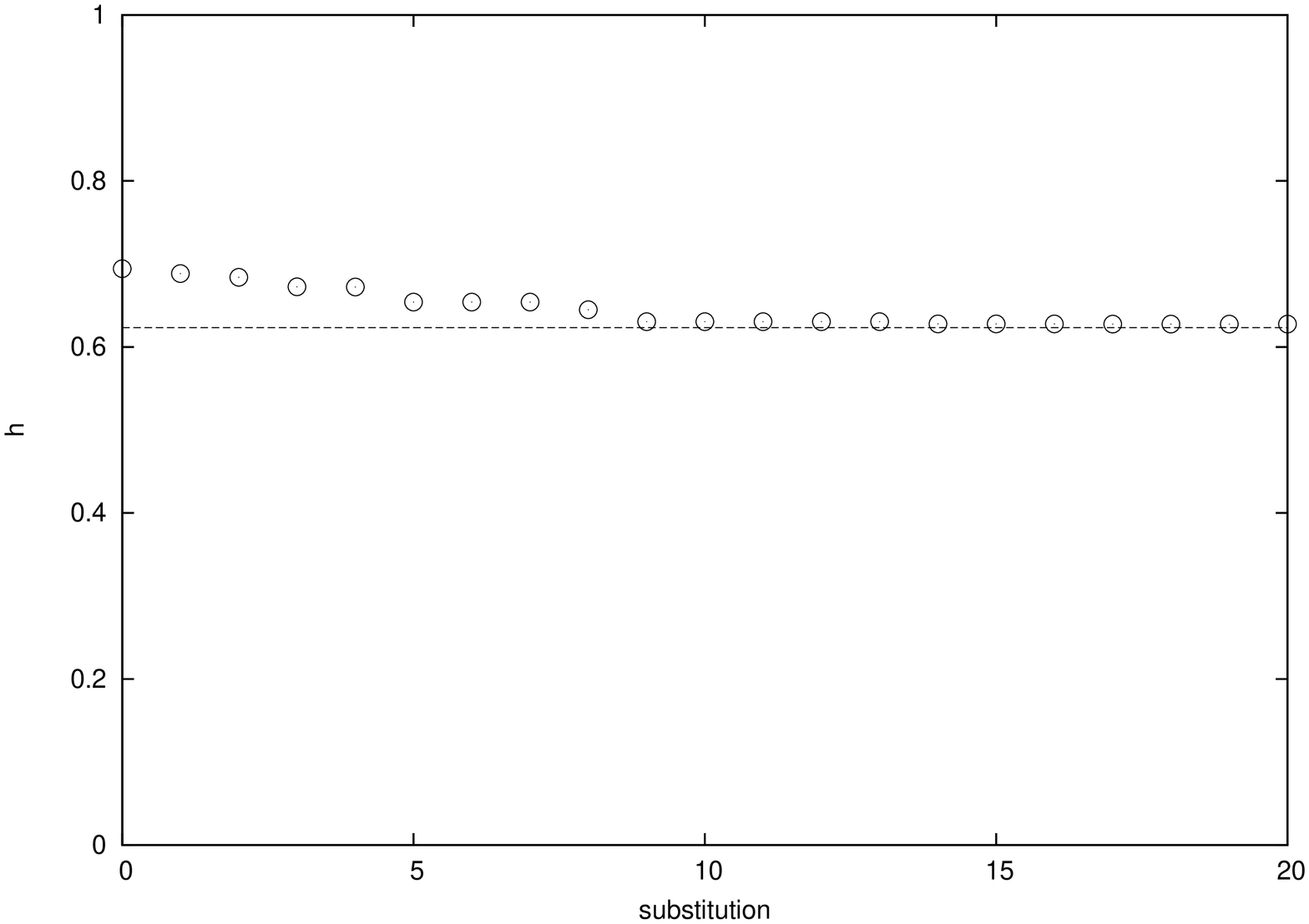}}}
\end{tabular}
\caption{\it Logistic map for $\lambda=3.8$ and entropy estimates by means of empirical frequencies, return times and NSRPS. The straight line corresponds to the Lyapunov exponent value.}
\label{figlogi}
\end{figure}

We took $\lambda = 4$, $3.8$ and $3.6$ (the graph of $\Lambda_{3.8}$ is shown in Figure \ref{figlogi} (map)). For all these three maps, the partition $\{ [0,\frac{1}{2}[,[\frac{1}{2},1] \}$ is generating (see \cite{buzzi}).

For $\lambda = 4$ there is a unique invariant measure, which is ergodic and absolutely continuous with respect to Lebesgue and whose density is $\rho (x) = \frac{1}{\pi \sqrt{x (1-x)}}$. Furthermore, the dynamical system $([0,1],\B ([0,1]),\rho (x) \textrm{d} x,\Lambda_4)$ is isomorphic to the shift on the Bernoulli process with alphabet $\{ 0,1 \}$ and parameter $\frac{1}{2}$. Thus, for the entropy it holds $h (\Lambda_4) = 1$.

About the maps $\Lambda_{3.8}$ and $\Lambda_{3.6}$ we remark that the assumptions of Theorem \ref{lyap} still hold and the dimension of the invariant measure is estimated to be very close to $1$ (see \cite{sprott}). Hence we assume to be reasonable to estimate the entropy by the Lyapunov exponent.

In Table \ref{table-logistic} we summarize the final entropy estimates obtained with the four methods for the three logistic maps, while in Figure \ref{figlogi} we show in graphical form the complete results for the map $\Lambda_{3.8}$.

% correzione (alcuni valori della tabella)
\begin{table}[!h]
\begin{equation*}
\renewcommand{\arraystretch}{1.5}  % aumenta la distanza predefinita tra le righe del fattore 1.5
\begin{array}{|l|r|r|r|r|}
\hline
\textrm{map}  & h_{\textrm{Lyap}} & h_{\textrm{EF}} & h_{\textrm{RT}} & h_{\textrm{NSRPS}} \ (N_{\textrm{sub}})\\ \hline
\Lambda_4     &            1.0000 &     0.997  &    0.959  &              1.000 \               (17)\\ \hline
\Lambda_{3.8} &            0.6234 &    0.652  &           0.610 &              0.628 \               (18)\\ \hline
\Lambda_{3.6} &            0.2646 &           0.348 &           0.314 &              0.269 \               (18)\\ \hline
\end{array}
\end{equation*}
\caption{\it Entropy estimates for the logistic maps $\Lambda_\lambda$. $N_{\textrm{sub}}$ is the number of pair substitutions executed when the stop condition \emph{StopCond} occurs.}
\label{table-logistic}
\end{table}

For the map $\Lambda_4$ the NSRPS method does not require any substitution to correctly estimate the entropy up to the sixth decimal digit. This is no surprise, since the symbolic process associated with $\Lambda_4$ is independent.

Instead, for the map $\Lambda_{3.8}$ it happens that, similarly to the NSRPS case of the map $L$ (see Figure~\ref{figlorenz}), there are pair substitutions which are more important than others in approximating the value of the entropy, as it can be noticed in Figure~\ref{figlogi}.

The entropy estimating algorithms give for the map $\Lambda_{3.6}$ results that are qualitatively similar to those of $\Lambda_{3.8}$. 

%--------------------------------------------
\subsubsection{Manneville-Pomeau maps}
Manneville maps exhibit dynamics %correzione
with long range correlations. They are defined by
\begin{equation*}
M_z x = x + x^z \ (\mathrm{mod} \ 1), \quad z \in \R^+.
\end{equation*}

\begin{figure}[!h] 
\centerline{\bf MANNEVILLE MAP}\vskip 10pt
\begin{tabular}{cc}
\fbox{map}&\fbox{EF}\\
{\raggedright{\includegraphics[height=6cm,angle=270]{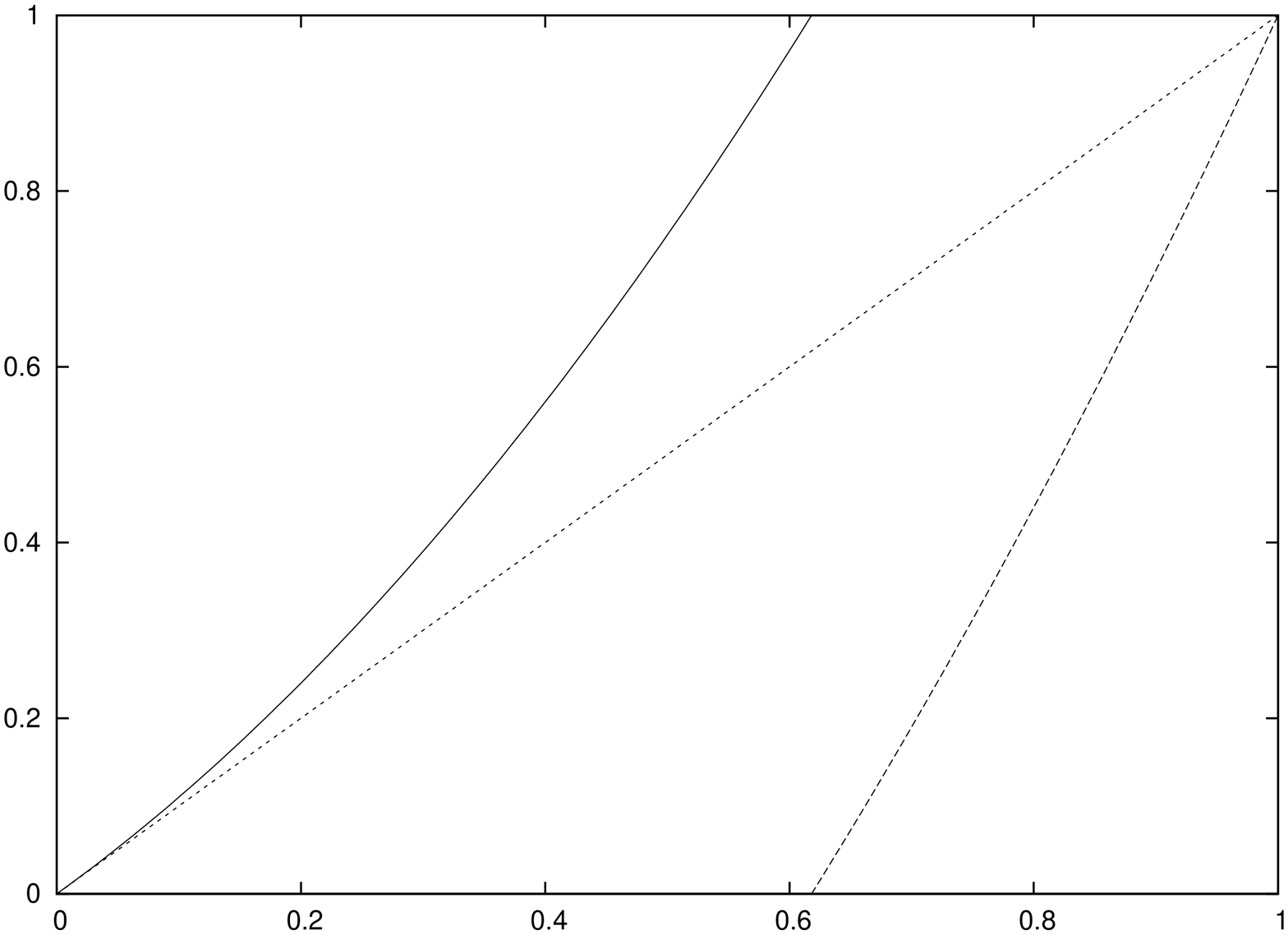}}} &
{\raggedleft{\includegraphics[height=6cm,angle=270]{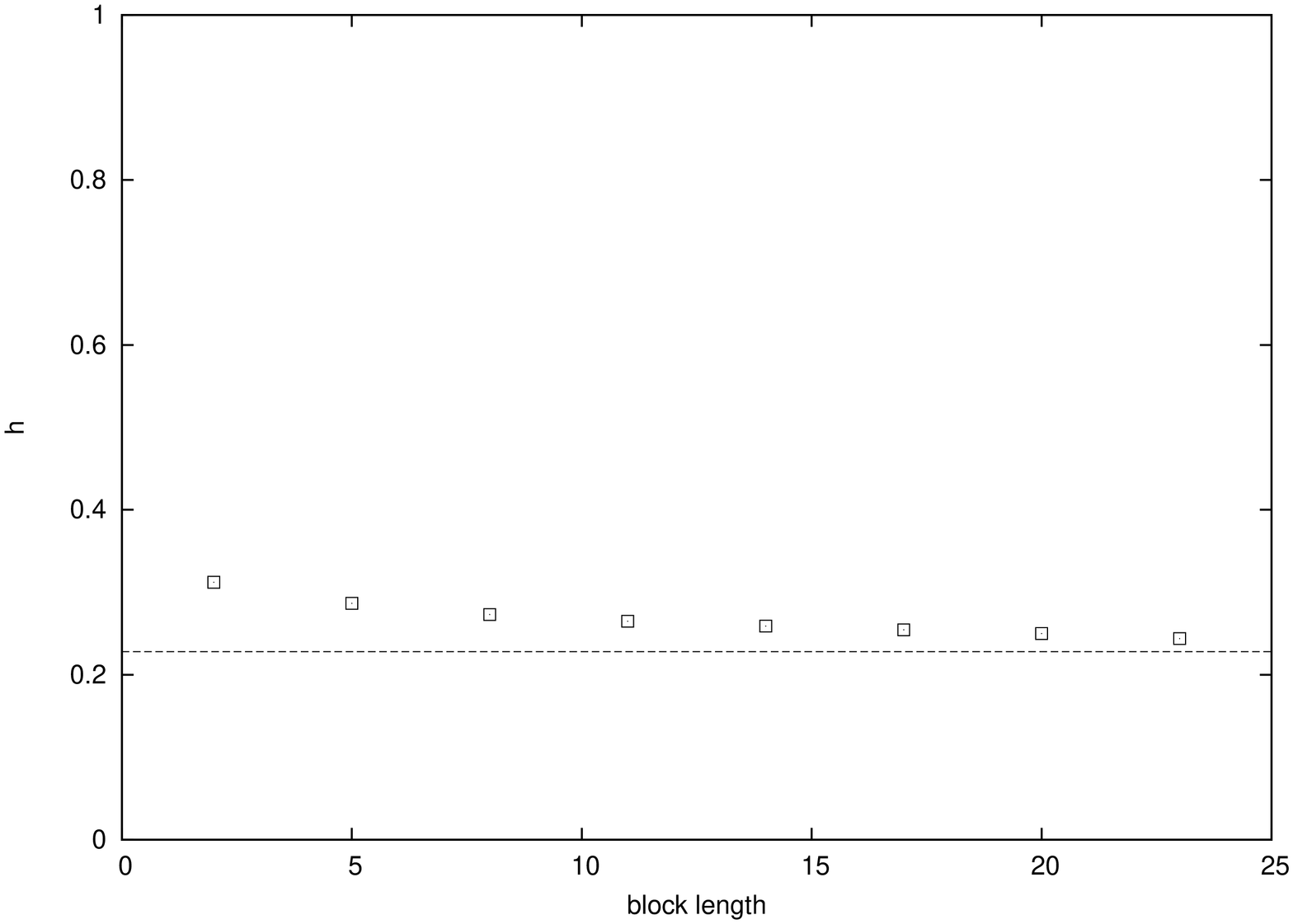}}}\\[125pt]
\fbox{RT}&\fbox{NSRPS}\\
{\raggedright{\includegraphics[height=6cm,angle=270]{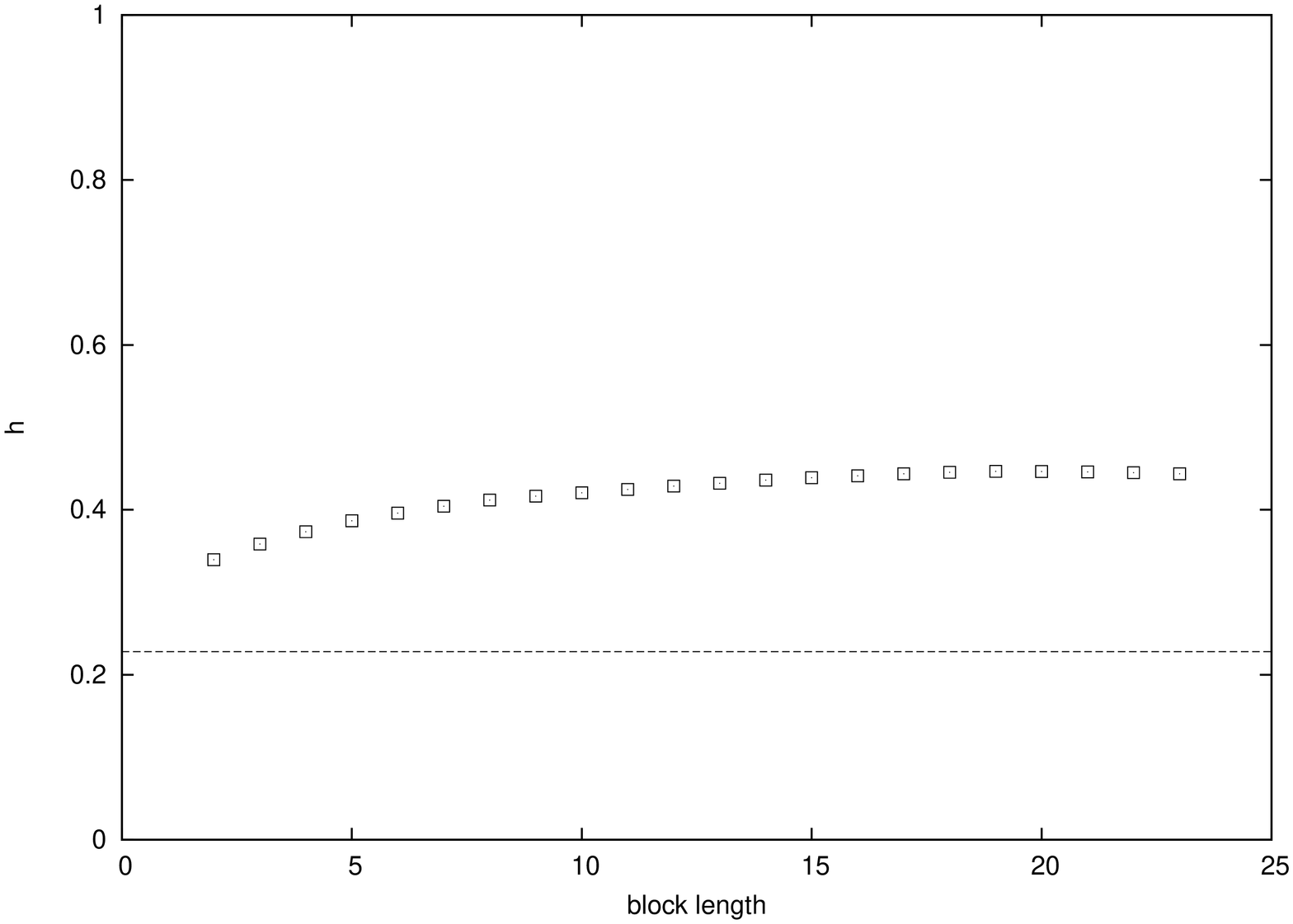}}}
&{\raggedleft{\includegraphics[height=6cm,angle=270]{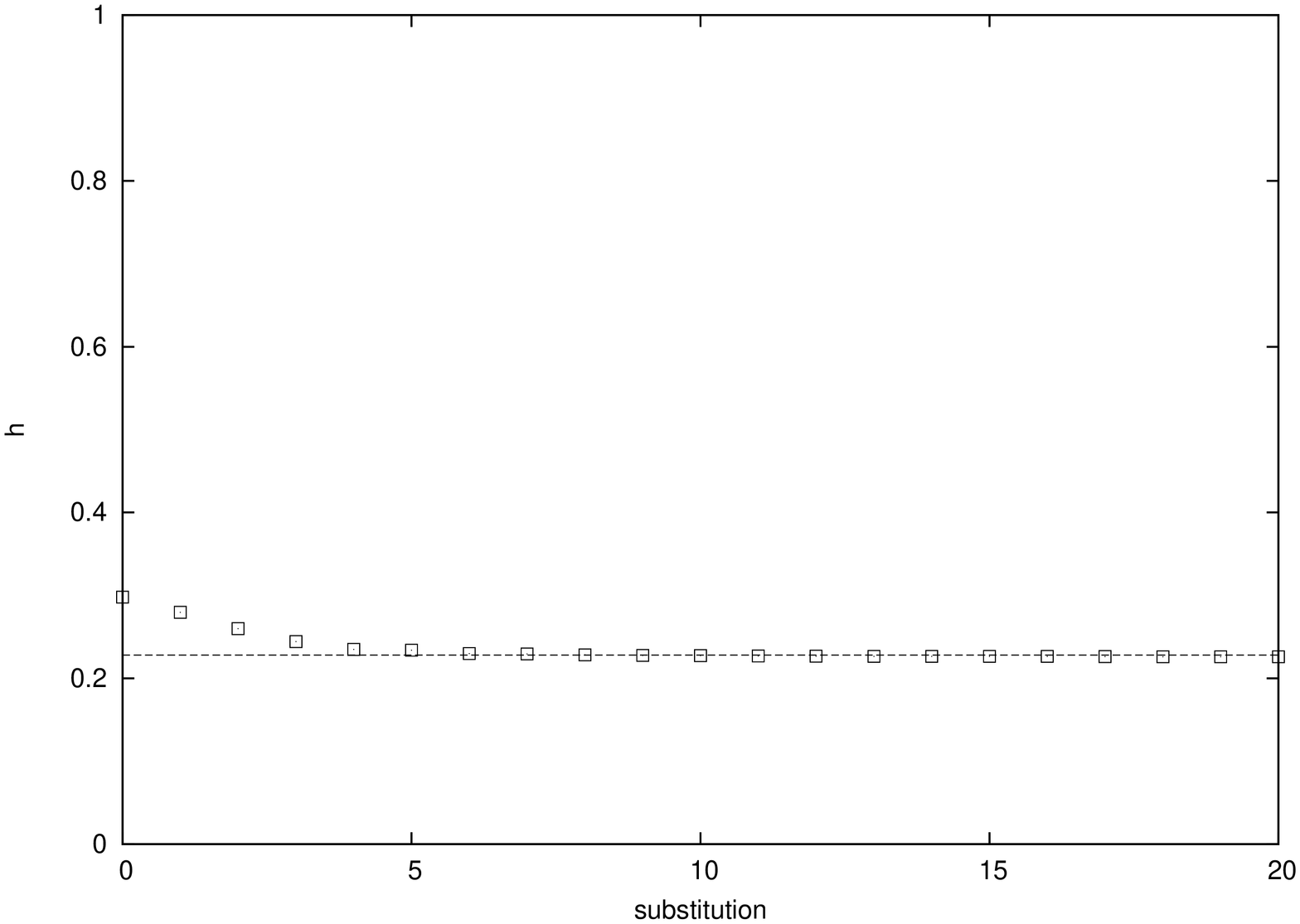}}}
\end{tabular}
\caption{\it Manneville map with $z = \frac{31}{16}$ and entropy estimates by means of empirical frequencies, return times and NSRPS. The straight line corresponds to the Lyapunov exponent value.}
\label{figmann}
\end{figure}

Such maps have great interest in physics and possess different characteristics as the exponent $z$ varies. We focused our attention on the values $1 < z < 2$, for which the maps admit a unique absolutely continuous invariant probability measure (with unbounded density).  For these parameters, the system has power law decay of correlations, and the rate is slower and slower as $z$ approaches $2$ (see \cite{viana}, section 3 e.g.). In this case the system has ``long memory'' and to estimate entropy by the empirical frequencies we would need long blocks. For $z \geq 2$ the absolutely continuous invariant measure is no longer finite.
We also remark that since those maps have bounded variation derivative,
in the cases where the absolutely continuous invariant measure is finite we can again estimate the entropy by the Lyapunov exponent.
We took values of $z$ which go very close to $2$: $z_1 = \frac{3}{2}$, $z_2 = \frac{7}{4}$, $z_3 = \frac{15}{8}$, $z_4 = \frac{31}{16}$, $z_5 = \frac{63}{32}$, $z_6 = \frac{127}{64}$ (see the plot of $M_{z_4}$ in Figure \ref{figmann}). For all $1 \leq i \leq 6$ it holds $M'_{z_i} (x) > 1$ for all $x \in ]0,1]$ and $M'_{z_i} (0^+) = 1$. For these maps the natural partitions $\{ [0,c_i[,[c_i,1] \}$, where $c_i \in ]0,1[$ is that value such that $M_{z_i} (c_i^-) = 1$ and $M_{z_i} (c_i^+) = 0$, are obviously generating.

The presence in $0$ of an indifferent fixed point is the main responsible for the peculiar behaviour of the Manneville maps. When, starting from a random point $x_0$, after a certain number $\overline{n}$ of iterations the point $M_z^{\overline{n}} x_0$ happens to be very close to $0$, the subsequent iterations remain very close to $0$ for a long time. This fact translates in having many consecutive zeros in the binary symbolic sequence associated with the orbit of $x_0$. These strings of zeros can be long even hundreds of thousands of bits or more. The closer to $2$ is the exponent $z$, the longer and more frequent these strings.

In carrying out the simulations for the Manneville maps and commenting their results, one cannot ignore the peculiarities of these maps. It turns out that the symbolic sequences we generated are too short to reflect the general characteristics of the maps. If in a sequence of $15$ millions bits there happen to be groups of consecutive zeros that are hundreds of thousands of bits long, then the results obtained from such a sequence cannot be completely reliable. The usual approach to this problem is considering many sequences, generated from different initial random points, and taking the averages of the estimates. For the map $M_{z_i}$ we considered $2 i$ sequences, with $1 \leq i \leq 6$. Still, the values obtained from the various sequences are quite different, so that we cannot consider completely reliable the averages as well.

Bearing in mind these considerations, we report in graphic form the results for the Manneville map $M_{z_4} = z + z^{\frac{31}{16}} \ (\mathrm{mod} \ 1)$ (see Figure \ref{figmann}), while the results for all the six Manneville maps considered are shown in Table \ref{table-manneville}.%correzione
%{\footnote{\bf The results of the maps $M_{z_i}$, $1 \leq i \leq 6$, are averages on the $2 i$ symbolic sequences considered.}}

% correzione (alcuni valori della tabella)
\begin{table}[!h]
\begin{equation*}
\renewcommand{\arraystretch}{1.5}  % aumenta la distanza predefinita tra le righe del fattore 1.5
\begin{array}{|l|r|r|r|r|}
\hline
\textrm{map}       & h_{\textrm{Lyap}} & h_{\textrm{EF}} & h_{\textrm{RT}} & h_{\textrm{NSRPS}} \ (N_{\textrm{sub}})\\ \hline
M_{\frac{3}{2}}    &             0.811 &     0.804  &     0.821  &              0.813 \      (18)\\ \hline
M_{\frac{7}{4}}    &             0.519 &     0.522  &     0.558  &        0.511  \      (20)\\ \hline
M_{\frac{15}{8}}   &             0.314 &    0.340 &    0.442 &              0.322 \      (19)\\ \hline
M_{\frac{31}{16}}  &             0.228 &    0.244  &    0.444  &        0.226  \      ( 21)\\ \hline
M_{\frac{63}{32}}  &             0.175 &     0.234 &     0.400  &        0.216  \      (21)\\ \hline
M_{\frac{127}{64}} &             0.168 &     0.214  &     0.358  &       0.196  \      (21)\\ \hline
\end{array}
\end{equation*}
\caption{\it Entropy estimates for the Manneville maps $M_{z_i}$. $N_{\textrm{sub}}$ is the average number of pair substitutions executed when the stop condition \emph{StopCond} occurs.}
\label{table-manneville}
\end{table}

For each Manneville map that we studied (except for $M_{\frac{7}{4}}$), the entropy estimates obtained through the NSRPS method were clearly the closest to the true entropy (which we assumed to be equal to the average Lyapunov exponent), although they were not as close as for the other maps or processes (see section \ref{sec:renewal}).

%-------------------------------------------- aggiunta
\subsubsection{A skew product}

We consider an example of a two dimensional system having long range 
correlations which is quite different from the Manneville map. Let us
consider the following map ${\cal S}:[0,1]^{2}\rightarrow \lbrack 0,1]^{2}$
defined by

\begin{equation}
{\cal S}(x,y)=(Ex,y+\alpha \phi (x)\ \textrm{mod } 1)
\end{equation}%
where: $\phi (x)=\left\{ 
\begin{array}{c}
1~if~x\geq \frac{1}{2} \\ 
0~if~x< \frac{1}{2}%
\end{array}%
\right. $, $\alpha $ is a diofantine irrational and $E$ is the one
dimensional piecewise expanding map considered in section \ref{pwexp}. In the system
the $x$ coordinate is subjected to a chaotic transformation, while the $y$
is rotated according to the value of $x$. Such systems preserve an
absolutely continuous invariant measure and are mixing. Some estimations for
the decay of correlations are given in \cite{Dol}.

We partitioned the unit square in four equal squares $Q_{1},...,Q_{4}$
having a common vertex at $(\frac{1}{2},\frac{1}{2})$. 

The entropy of ${\cal S}$ with respect to the partition $\{Q_{1},...,Q_{4}\}$
is the same as the entropy of $E$, indeed the rotation has zero entropy and
a symbolic orbit for the two dimensional system can be constructed by the
information given by its symbolic orbit for the one dimensional map $E$ and
the information relative to the rotation part. Although the entropy is the
same, its estimation is much more complicated, as the experiments show. 

\begin{figure}[!h]
\centerline{\bf SKEW PRODUCT}\vskip 10pt 
\begin{tabular}{cc}
\fbox{EF}&\fbox{RT}\\ 
{\raggedright{\includegraphics[height=6cm,angle=270]{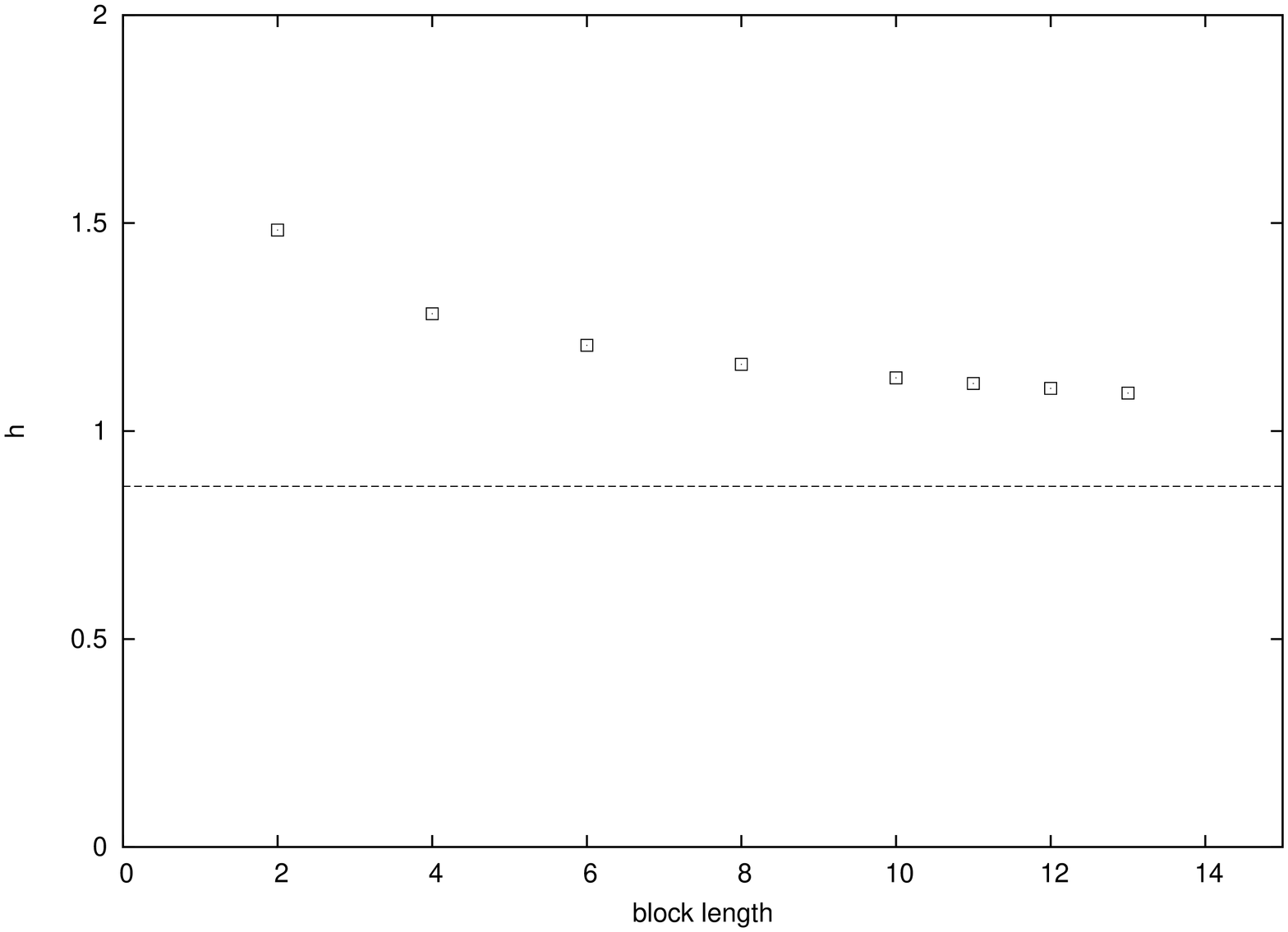}}}
&{\raggedleft{\includegraphics[height=6cm,angle=270]{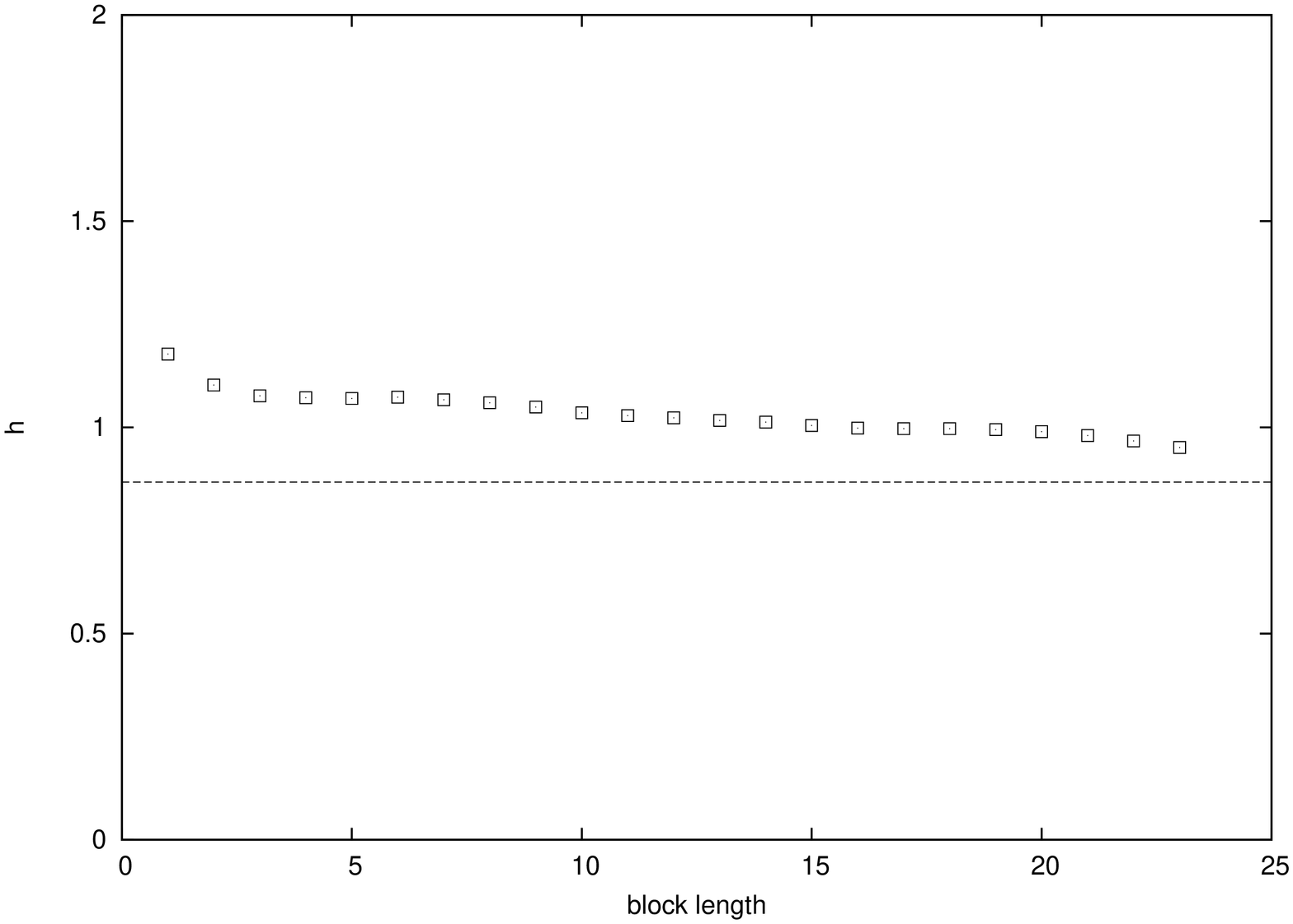}}}
\end{tabular}\\[15pt] 
\centerline{\fbox{NSRPS}}
\centerline{\includegraphics[height=6cm,angle=270]{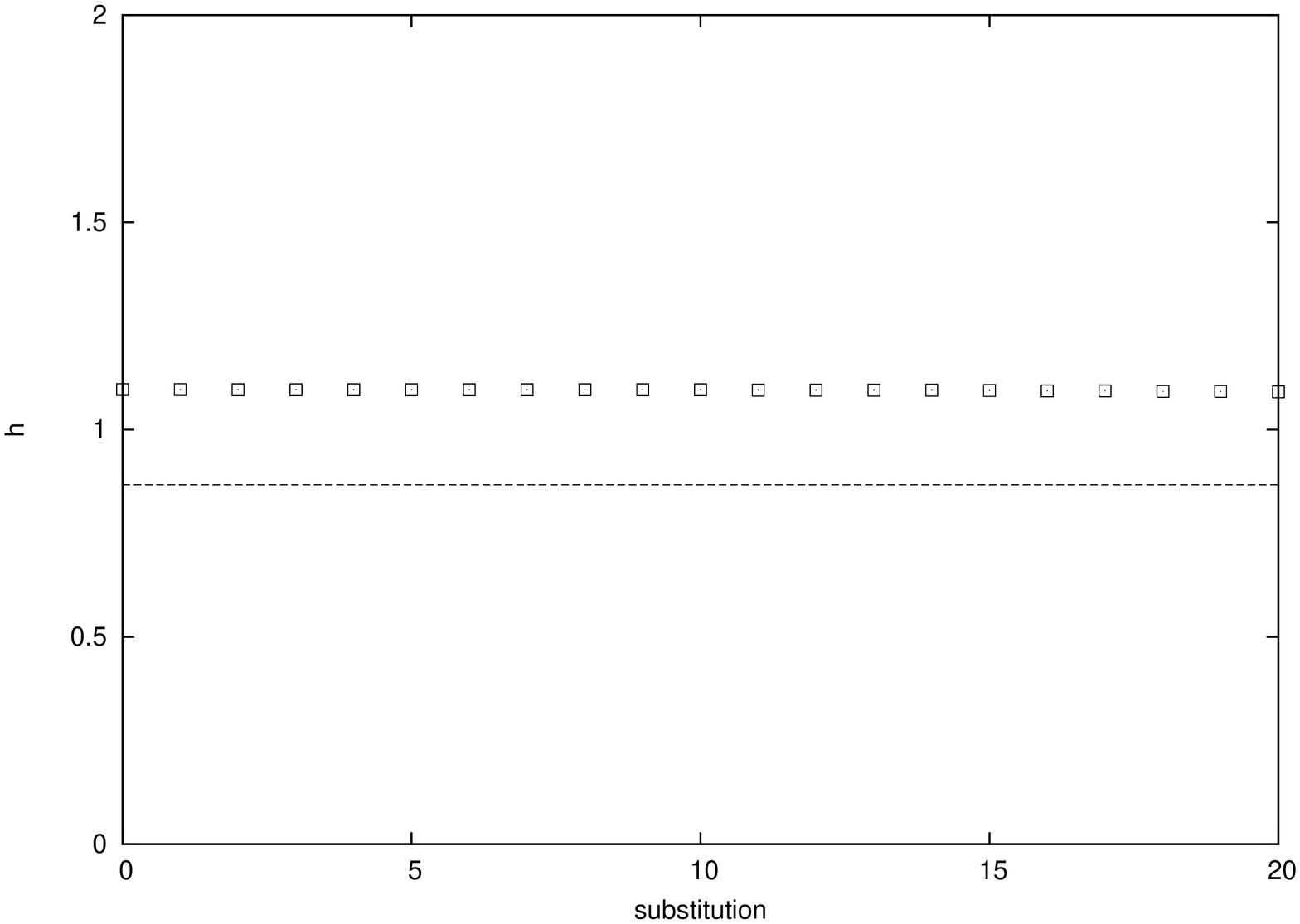}}
\caption{\it Results for the skew product ${\cal S}$: entropy estimates by means of empirical frequencies, return times and NSRPS. The straight line corresponds to the entropy value.}
\label{figmb} 
\end{figure}

In figure \ref{figmb} we consider the case where $\alpha =\frac {1+\sqrt{5}}{2}$ is the golden ratio.
The empirical frequencies and the substitutions seem to converge to a value
which is slightly greater than the true entropy. The return time instead seems to
better approximate the entropy in this case.

%_____________________________

%--------------------------------------------
\subsection{Renewal processes} \label{sec:renewal}
Apart from the symbolic sequences obtained from ergodic transformations of the unit interval, we considered sequences taken from the so-called renewal processes.

A renewal process is a stationary process with alphabet $\{ 0,1 \}$ for which the distances between consecutive ones are independent and identically distributed random variables. When a symbol `1' occurs, the sequence forgets all its past and the probability of having the next `1' after $j$ bits is $p_j$, where $0 \leq p_j \leq 1$ and $\sum_{j=1}^\infty p_j = 1$.

We considered such renewal processes, with $p_1 = p_2 = \ldots = p_{2^k} = \frac{1}{2^k}$, $5 \leq k \leq 9$, which we shall indicate with $RP_{2^k}$.

For these renewal processes, the value of their entropy can be calculated exactly. We recall in fact that the entropy of a process is the number of bits per symbol that are necessary to describe the process itself. The quantity $C = - \sum_j p_j \log_2 p_j$ represents the number of bits that one needs to describe the process of the jumps between consecutive ones. In other words, $C$ is the entropy of a random variable which describes the length of the jumps. If $n$ is large, with $n$ jumps ($nC$ bits) we describe a sequence long about $n \overline{L}$ symbols, where $\overline{L}$ is the average length of the jumps. Thus,
\begin{equation*}
h (RP_{2^k}) \approxlim \frac{n C}{n \overline{L}} = \frac{- \sum_{j \geq 1} p_j \log_2 p_j}{\sum_{j \geq 1} j p_j}.
\end{equation*}
In our cases, where $p_1 = \ldots = p_{2^k} = \frac{1}{2^k}$ and $p_j = 0$ for $j > 2^k$, we have
\begin{equation*}
h (RP_{2^k}) = \frac{2 k}{2^k + 1}.
\end{equation*}

\begin{figure}[!h]
\centerline{\bf RENEWAL PROCESS}\vskip 10pt 
\begin{tabular}{cc}
\fbox{EF}&\fbox{RT}\\ 
{\raggedright{\includegraphics[height=6cm,angle=270]{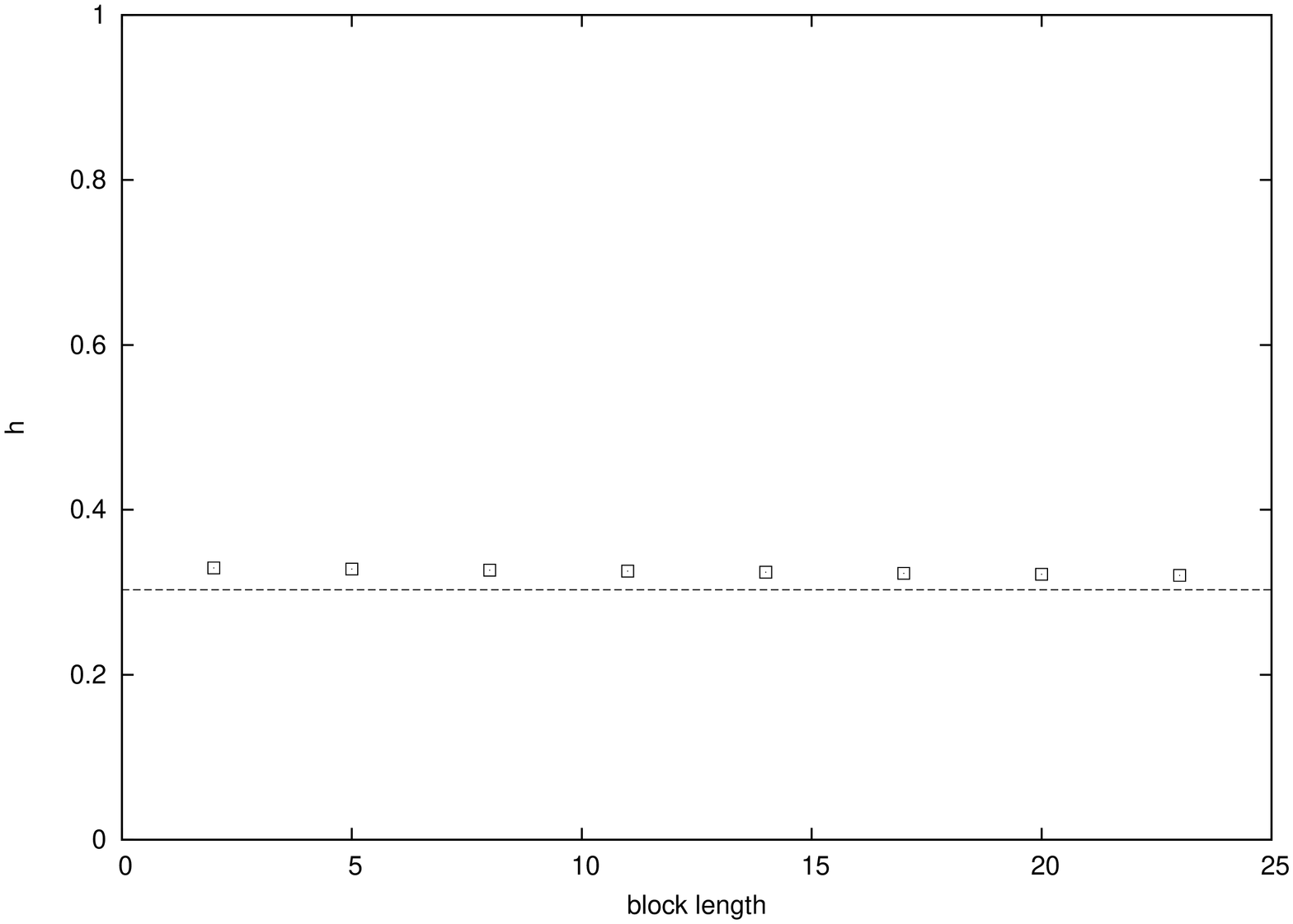}}}
&{\raggedleft{\includegraphics[height=6cm,angle=270]{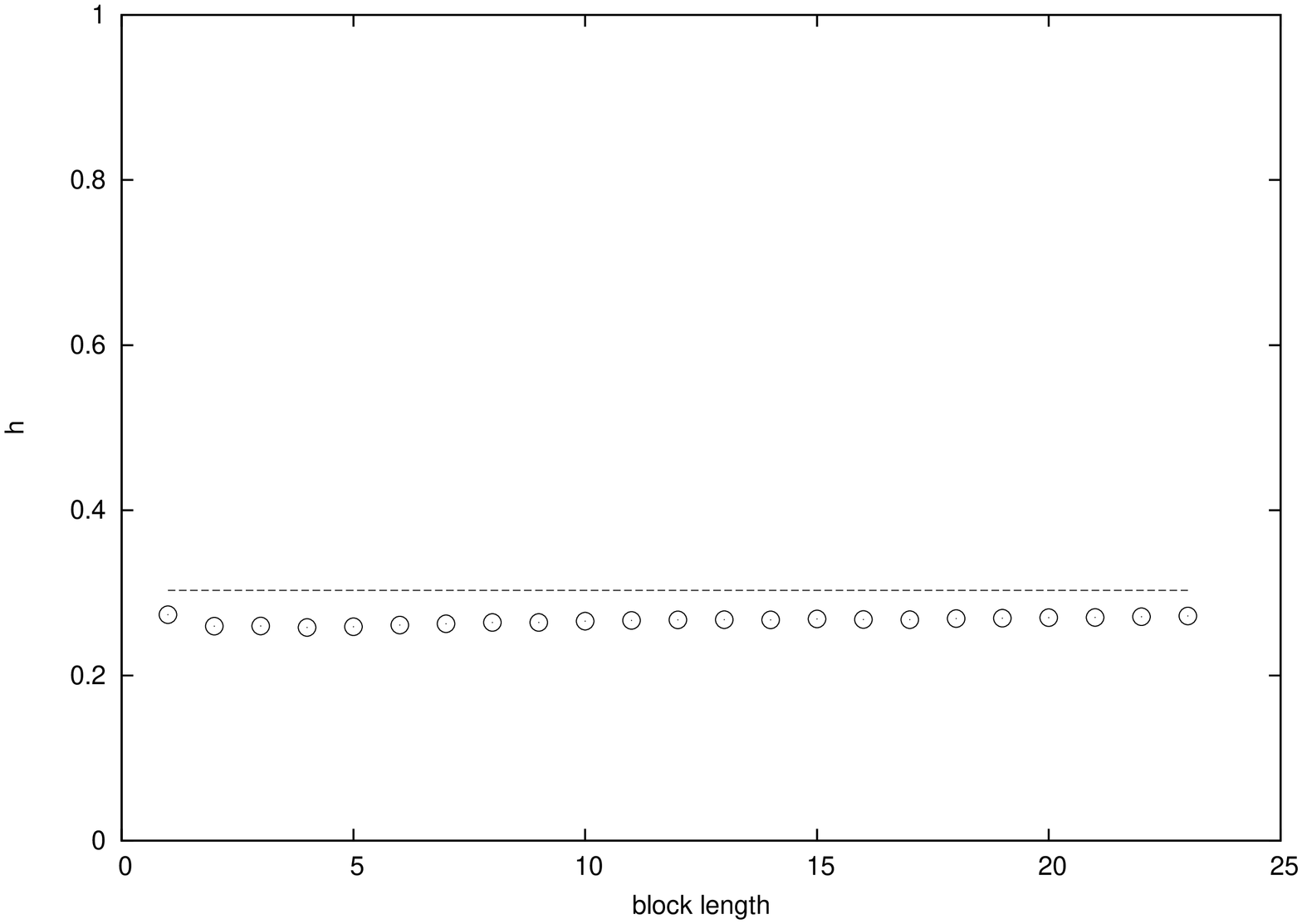}}}
\end{tabular}\\[15pt] 
\centerline{\fbox{NSRPS}}
\centerline{\includegraphics[height=6cm,angle=270]{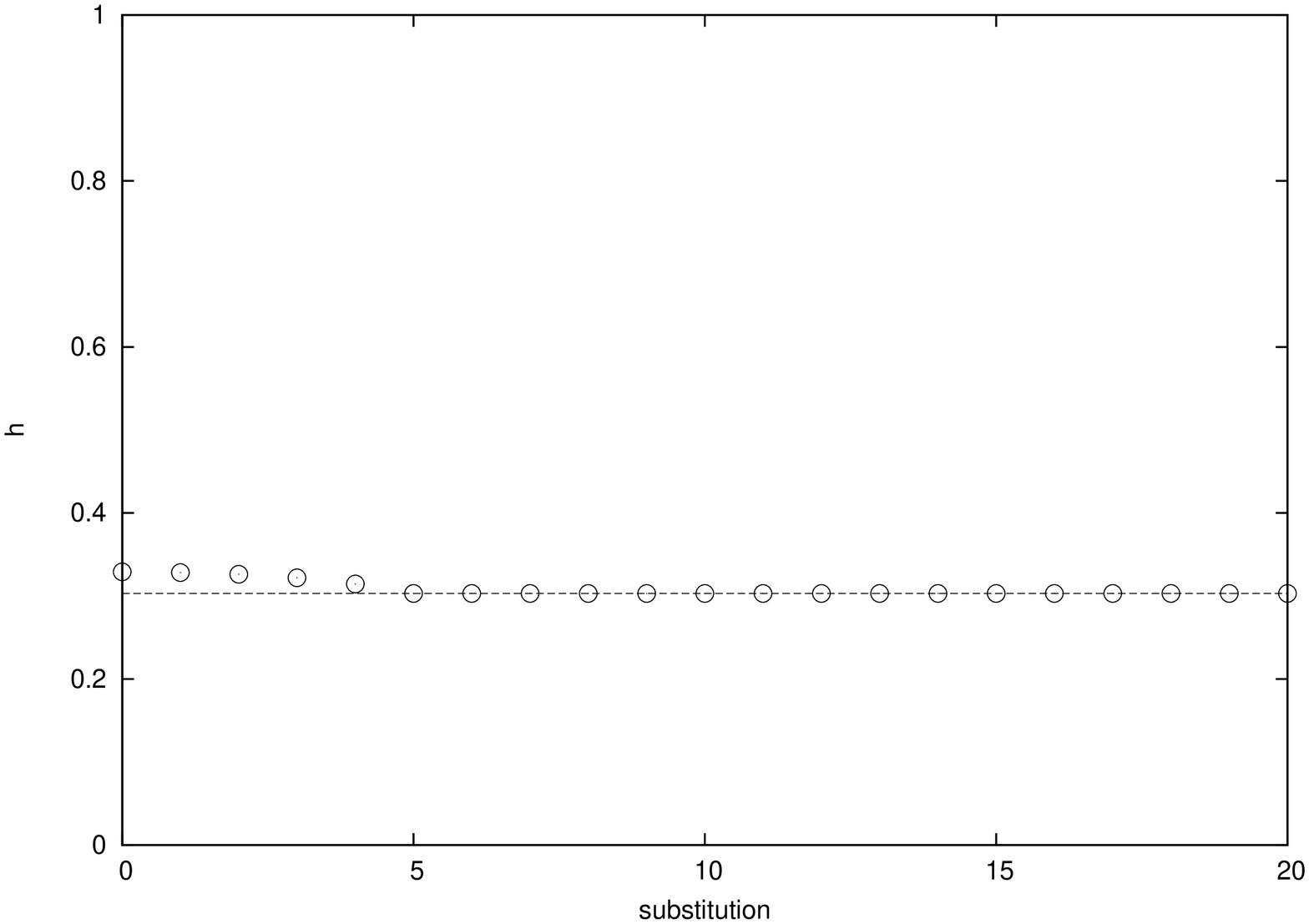}}
\caption{\it Results for the renewal process $RP_{32}$: entropy estimates by means of empirical frequencies, return times and NSRPS. The straight line corresponds to the entropy value.}
\label{figrp} 
\end{figure}

In Table \ref{table-renewal} we show the results of the entropy estimates for the renewal processes $RP_{2^k}$ and those of $RP_{32}$ are also plotted in Figure \ref{figrp}.

\begin{table}[!h]
\begin{equation*}
\renewcommand{\arraystretch}{1.5}  % aumenta la distanza predefinita tra le righe del fattore 1.5
\begin{array}{|l|r|r|r|r|}
\hline
\textrm{map} &     h    & h_{\textrm{EF}} & h_{\textrm{RT}} & h_{\textrm{NSRPS}} \ (N_{\textrm{sub}})\\ \hline
RP_{32}      & 0.303030 &           0.320 &           0.272 &           0.303067 \               (11)\\ \hline
RP_{64}      & 0.184615 &           0.196 &           0.153 &           0.184793 \               (22)\\ \hline
RP_{128}     & 0.108527 &           0.115 &           0.110 &           0.108498 \               (25)\\ \hline
RP_{256}     & 0.062257 &           0.066 &           0.055 &           0.062239 \               (18)\\ \hline
RP_{512}     & 0.035088 &           0.037 &           0.039 &           0.035112 \               (16)\\ \hline
\end{array}
\end{equation*}
\caption{\it Entropy estimates for the renewal processes $RP_{2^k}$. $N_{\textrm{sub}}$ is the number of pair substitutions executed when the stop condition \emph{StopCond} occurs.}
\label{table-renewal}
\end{table}

 For this process, the substitutions method gives an excellent approximation of the entropy already after five pair substitutions. After these substitutions all the memory of the process has been transferred to the distribution of the pairs, so that the sequence has become $1$-Markov.

\section{Conclusions and final remarks} \label{sec:conclusions}
\begin{figure}[h]
\begin{center}
\includegraphics[height=10cm,angle=270]{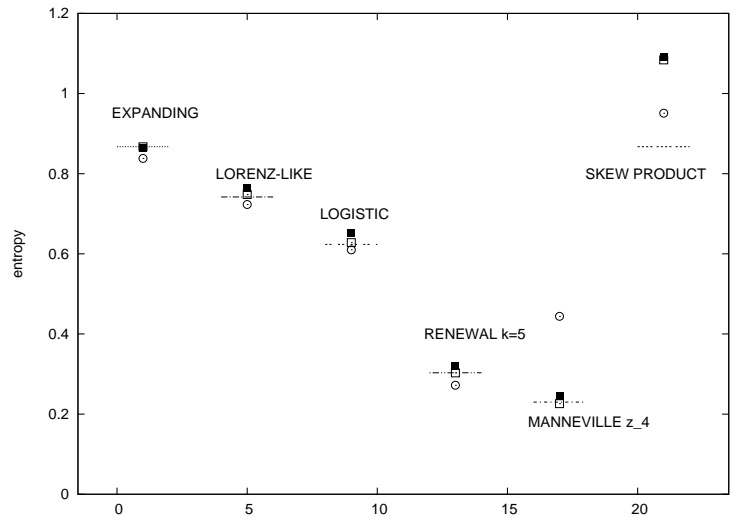}
\caption{\it Entropy estimates for the maps $\Lambda_{3.8}$, $E$, $L$, $M_{\frac{31}{16}}$ and the renewal process $RP_{32}$: symbol empty $\square$ refers to the NSRPS value under \emph{StopCond} condition; full $\square$ refers to the EF value and $\circ$ refers to the RT method. Straight lines show the entropy values (for the maps they are the estimated Lyapunov exponents).}
\label{fig:finale}
\end{center}
\end{figure}
The performance of the three symbolic methods is summarized in Figure \ref{fig:finale}. 
 Summarizing, NSRPS results to be the method that best approximates the entropy value. To this aim, it is a fast and light computational tool that may be used also for systems having low entropy 
or long range correlations where other statistical methods fail.

This paper shows for the first time a comparison in entropy estimation among NSRPS and other well-known methods. The results also open some further questions about NSRPS:
\begin{itemize}
\item how to prove an analogous of Theorem \ref{empirical_frequencies} for NSRPS giving a sufficient number of substitutions in function of the length of the string?

\item are there other meaningful substitution methods (different from the recipe given in Theorem \ref{suff_cond}) that may be proved to be (at least) sufficient for Theorem \ref{main_theorem} to hold?

\item can the joint use of NSRPS and Lyapunov exponent (which are both fast converging and fastly computable) together with Theorem \ref{lyap} give a particularly good method to numerically estimate the
Hausdorff dimension of an attractor?

\item concerning the applications of NSRPS to non-artificial processes, such as literary texts, biological sequences (DNA, proteins) and time series in general, what interesting features of the driving dynamics may be extracted?

\item NSRPS method might be the core of some data compression algorithm (see \cite{moffat}). This should pave the way to some investigations towards its compression capabilities in comparison with other well-known algorithms. We remark that data compression procedures have also been successfully used as entropy estimators (see e.g. \cite{licatone} and \cite{kont}).
\end{itemize}

%---------------------------------------------------------------------------------------
\noindent{\bf{Acknowledgements. }}The work of G.M. was supported by a post-doc research scholarship ``Compagnia di San Paolo'' awarded by the Istituto Nazionale di Alta Matematica ``F. Severi''.

\end{document}